\documentclass{IEEEtran}

\usepackage{amsfonts}
\usepackage[cmex10]{amsmath}
\allowdisplaybreaks[1]  
\usepackage{amsthm}
\usepackage{amssymb}
\usepackage{mdwmath}
\usepackage{mathtools}
\usepackage{cases}
\usepackage{empheq}

\theoremstyle{definition}

\usepackage{tabularx}
\usepackage{array}
\usepackage{mdwtab}
\usepackage{multirow}
\usepackage{booktabs}  
\usepackage{diagbox} 
\usepackage{makecell}

\usepackage{textcomp}
\def\BibTeX{{\rm B\kern-.05em{\sc i\kern-.025em b}\kern-.08em
T\kern-.1667em\lower.7ex\hbox{E}\kern-.125emX}}

\usepackage[pdftex]{graphicx}
\usepackage{caption}  
\usepackage[labelformat=simple]{subcaption}
\graphicspath{ {./img/} }

\usepackage{cite}

\usepackage{makeidx}

\usepackage{algorithm}
\usepackage{algpseudocode}
\algrenewcommand\algorithmicindent{1em}

\newcommand{\ParState}[1]{\State \parbox[t]{\dimexpr\linewidth-\algorithmicindent}{#1\strut}}  

\usepackage{hyperref} 
\usepackage{cleveref} 

\hypersetup{colorlinks = true, 
	    linkcolor = black, 
	    urlcolor = black,
        citecolor = black} 

\begin{document}
\bstctlcite{IEEEexample:BSTcontrol}  

\title{Multi-View Wireless Sensing via Conditional Generative Learning: Framework and Model Design}

\author{
\IEEEauthorblockN{
    Ziqing~Xing, 
    Zhaoyang~Zhang, 
    Zirui~Chen, 
    Hongning~Ruan, 
    Zhaohui~Yang, 
    and Zhiyong Feng
    \vspace{-0.7cm}
}
\thanks{Part of this work has been presented at 2025 IEEE International Symposium on Personal, Indoor and Mobile Radio Communications (PIMRC) \cite{xing2025conference}.}
\thanks{This work was supported in part by National Natural Science Foundation of China under Grants 62394292, Zhejiang Provincial Key R\&D Program under Grant 2023C01021, and the Fundamental Research Funds for the Central Universities under Grant 226-2024-00069. 
(\textit{Corresponding author: Zhaoyang Zhang})}
\thanks{Z.~Xing and Z.~Zhang are with the College of Information Science and Electronic Engineering, Zhejiang University, Hangzhou 310027, China, 
also with the Institute of Fundamental and Transdisciplinary Research, Zhejiang University, Hangzhou 310058, China, 
and also with the Zhejiang Provincial Laboratory of Multi-Modal Communication Networks and Intelligent Information Processing, Hangzhou 310027, China 
(e-mail: ziqing\_xing@zju.edu.cn; ning\_ming@zju.edu.cn).}
\thanks{Z.~Chen, H.~Ruan and Z.~Yang are with the College of Information Science and Electronic Engineering, Zhejiang University, Hangzhou 310027, China, 
and also with the Zhejiang Provincial Laboratory of Multi-Modal Communication Networks and Intelligent Information Processing, Hangzhou 310027, China
(e-mail: ziruichen@zju.edu.cn; rhohenning@zju.edu.cn; yang\_zhaohui@zju.edu.cn).}
\thanks{Zhiyong Feng is with the Key Laboratory of Universal Wireless Communications, Ministry of Education, and also with the
School of Information and Communication Engineering, Beijing University of Posts and Telecommunications, Beijing, 100876, China
(email: fengzy@bupt.edu.cn).}
}

\maketitle

\begin{abstract}    
    In this paper, we incorporate physical knowledge into learning-based high-precision target sensing using the multi-view channel state information (CSI) between multiple base stations (BSs) and user equipment (UEs). Such kind of multi-view sensing problem can be naturally cast into a conditional generation framework. To this end, we design a bipartite neural network architecture, the first part of which uses an elaborately designed encoder to fuse the latent target features embedded in the multi-view CSI, and then the second uses them as conditioning inputs of a powerful generative model to guide the target's reconstruction. Specifically, the encoder is designed to capture the physical correlation between the CSI and the target, and also be adaptive to the numbers and positions of BS-UE pairs. Therein the view-specific nature of CSI is assimilated by introducing a spatial positional embedding scheme, which exploits the structure of electromagnetic(EM)-wave propagation channels. Finally, a conditional diffusion model with a weighted loss is employed to generate the target's point cloud from the fused features. Extensive numerical results demonstrate that the proposed generative multi-view (Gen-MV) sensing framework exhibits excellent flexibility and significant performance improvement on the reconstruction quality of target's shape and EM properties.
\end{abstract}

\begin{IEEEkeywords}
Integrated sensing and communication (ISAC), multi-view sensing, generative model, diffusion model
\end{IEEEkeywords}

\section{Introduction} \label{sec:introduction}
The sixth-generation (6G) wireless network will serve as a foundational service for various emerging applications, such as autonomous driving, extended reality (XR), intelligent robotics, etc \cite{chen2024wBAIM}.   
These applications not only demand reliable wireless communication but also require the acquisition of environmental information. 
Integrated sensing and communication (ISAC) is a promising solution to address these requirements efficiently, 
aiming to reuse wireless communication systems to enable environment sensing functions, including detection, localization, and imaging \cite{liu2022integrated}.   
Due to ISAC's efficiency in spectrum utilization and hardware resources, it has emerged as a focal point of current 6G research.

The fundamental issues in mono-static and bi-static ISAC scenarios have been widely studied, including beamforming strategy \cite{zhuo2022multibeam}, hardware architecture \cite{zhang2018multibeam}, performance trade-off \cite{xiong2023fundamental}, etc. 
Building upon this foundation, some research efforts have explored emerging applications within these fundamental ISAC scenarios. 
For example, \cite{che2024novel} leverages the initial access to achieve environment sensing and user localization, while \cite{jiang2024EMpropertySensing} proposes an electromagnetic (EM) property sensing scheme for scatterers. 
However, the link between a single transceiver pair only captures partial environmental information, 
making it difficult to achieve high-quality sensing results from such a limited observation \cite{tong2023multi}. 

{
Given the limitations of single-link sensing, a series of studies \cite{xing2023physics, cui2025blockage, zhou2024multi} drew inspiration from multi-static radar, 
leveraging multi-view observations formed by distributed transceivers to enhance the system's sensing capability. 
This approach introduce additional spatial diversity, which helps to address challenges such as non-line-of-sight sensing and occlusion effects \cite{tong2022environment}. 
However, existing multi-view sensing schemes are primarily based on conventional radar models and traditional signal processing methods,  
which typically use radar cross-section (RCS) to approximate the macroscopic scattering characteristics of targets, 
and perform target parameter estimation based on simplified EM propagation models and target statistical priors \cite{yu2021efficient}. 
\cite{jiang2025multi-BS} applied rigorous EM scattering principles to model a multi-view ISAC problem, but it still only utilized the target's sparsity to construct the iterative inversion algorithm. 
Therefore, the effectiveness of traditional multi-view sensing methods is limited by the accuracy of the channel model and the statistical prior of targets. 
}

To overcome the reliance of traditional algorithms on statistical priors and explicit forward modeling, a number of research efforts have explored the application of artificial intelligence (AI) to address challenges in ISAC systems, such as hardware impairment compensation \cite{mateos2025modelbaseEnd2End}, intelligent beamforming \cite{qi2024deep}, and high-precision parameter estimation \cite{naoumi2024AoA_AoD}. 
Furthermore, \cite{liu2023OAC-MVpooling} proposed a distributed aggregation scheme for multi-device sensing features to perform target classification. 
These advancements demonstrate the remarkable capability of AI in enabling efficient ISAC. 
In addition, environmental sensing is essentially an inverse problem based on the EM propagation laws, 
thus combining model-driven and data-driven approaches is a promising direction toward more fine-grained sensing. 
Currently, researchers have designed physics-informed neural networks to solve the EM inverse scattering problem, a fundamental physical problem in wireless imaging \cite{xing2025vbim,wang2023MSDLS,ruan2025NeuralSurf}. 
However, these methods are not specifically tailored for wireless communication systems and cannot effectively handle diverse scenario configurations. 

Different from classical AI techniques that build deterministic mappings, generative AI (GenAI) leverages probabilistic models to infer latent distributions from large-scale data, offering a more versatile way to discover fundamental mechanisms \cite{chen2024towards}. 
Recently, GenAI has inspired novel approaches for radio frequency signal processing \cite{chi2024RFdiff}, channel representation \cite{chen2025CD}, and intelligent localization \cite{chen2025AL}. 
These advancements demonstrate GenAI’s immense potential in uncovering the underlying physical laws of wireless data and performing rule-based generation for both signals and scenario characteristics, propelling a new path toward intelligent ISAC systems. 
{\color{black}Wang et al. outlined potential GenAI-enhanced ISAC physical layer technologies \cite{wang2024genISAC}, and further proposed two diffusion model-assisted case studies: a human flow detection scheme \cite{wang2024genHumanFlow} and a secure sensing system \cite{wang2025genSecureISAC}. }
Jiang et al. further employed a diffusion model for the target imaging task in a mono-static ISAC scenario \cite{jiang2024EMdiff}. 
However, these existing works mainly apply GenAI to partial processing stages of sensing algorithms and lack scalability for multi-BS multi-UE scenarios. 
These limitations constrain the effectiveness of GenAI in multi-device collaborative sensing systems. 

To address these issues, this paper proposes a novel generative multi-view (Gen-MV) sensing framework, 
which fuses the channel state information (CSI) from multiple BSs and UEs to achieve EM imaging of the target within the region of interest (RoI). 
In response to variable view configurations in sensing scenarios, we introduce a general multi-view channel encoder for target feature extraction. 
Its main function is to eliminate target-irrelevant BS and UE positions  from the CSI under different views, and fuse the multi-view channel features to integrate scene information. 
Subsequently, we employ a point cloud diffusion model with weighted loss to reconstruct the target shape and EM properties under the guidance of the extracted target features. 
The proposed Gen-MV sensing framework embeds physics-informed models into conditional generative learning, effectively handling dynamic changes in the positions and quantities of ISAC devices, and enhancing sensing quality with multi-view channels. 
The main contributions of this paper are summarized as follows: 
\begin{itemize}
    \item Based on the EM scattering channel model, we formulate the multi-view sensing problem using uplink CSI, explore the advantages of multi-view joint processing, and propose a Gen-MV sensing framework as a general solution. 
    It mainly consists of a multi-view channel encoder and a target generator with conditioning mechanisms. 
    \item In the multi-view channel encoder, we decouple the channel features from the positions of the BS and UE through a multiplicative positional embedding. 
    Then, we implement three baseline encoders using classical architectures and further design an interleaved correlation learning architecture by leveraging the intrinsic physical structure of the multi-view channel.
    \item We represent the target using a shape-EM point cloud and employ a diffusion model to achieve generative target reconstruction. 
    To address the distributional differences in the geometric shape and EM properties of the target, we propose a shape-EM weighted loss function to train the complete model.
    \item Through extensive numerical experiments, we demonstrate that the proposed schemes can flexibly accommodate variable positions and quantities of BSs and UEs, leveraging multi-view CSI to achieve effective target imaging. 
    Ablation studies further validate the working mechanisms and design rationale of our solutions. 
\end{itemize}

The remainder of this paper is organized as follows. 
Section \ref{sec:system_model} introduces the system model, including the scenario setup and EM channel modeling. 
In Section \ref{sec:MV_gen}, we propose the Gen-MV sensing framework, 
introduce the design of the multi-view channel encoder, 
and derive the target reconstruction process based on the diffusion model. 
Section \ref{sec:numerical_results} provides performance evaluations and ablation studies of our proposed scheme. 
Finally, we conclude our work in Section \ref{sec:conclusion}. 

\textit{Notations}: 
Fonts $a$, $\boldsymbol{a}$, and $\mathbf{A}$ represent scalars, vectors, and matrices, respectively. 
$\boldsymbol{a}[m]$ and $\mathbf{A}[m, n]$ denote the $m$-th element of vector $\boldsymbol{a}$ and the $(m, n)$-th element of matrix $\mathbf{A}$, respectively.
$\mathrm{vec}(\mathbf{A})$ denotes the vectorization of matrix $\mathbf{A}$ and $\mathrm{reshape}(\boldsymbol{a}, (M, N))$ denotes reshaping vector $\boldsymbol{a}$ into an $M\times N$ matrix. 
$\otimes$ denotes the Kronecker product and $\ast$ denotes the Khatri–Rao product. 
$\odot$, $\oplus$, and $\nabla$ denote the Hadamard product, concatenation, and gradient operators, respectively. 
$D_{\mathrm{KL}}\left(q(\boldsymbol{x}) \parallel p(\boldsymbol{x})\right)$ denotes the Kullback–Leibler (KL) divergence from distribution $q(\boldsymbol{x})$ to distribution $p(\boldsymbol{x})$. 
In physics, $\varepsilon_0$, $\mu_0$, and $c$ represent the permittivity, permeability, and speed of light in vacuum, respectively. 

\begin{figure} [!t] 
\centering
\hspace{0.3cm}
\includegraphics[width=0.86\linewidth]{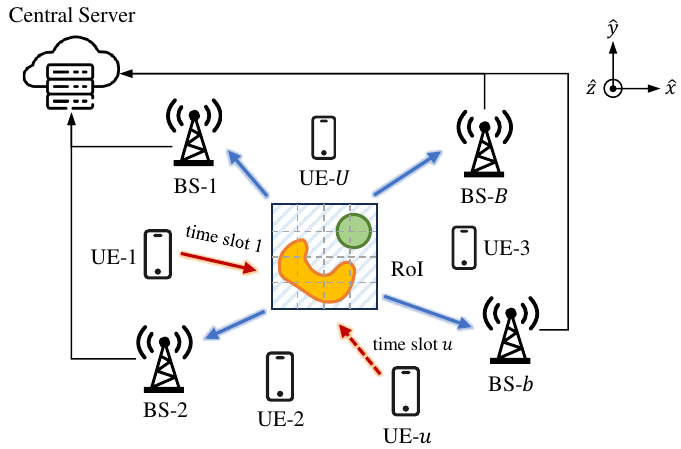}
\vspace{-0.1 cm}
\caption{The considered multi-view uplink sensing scenario.}
\label{fig:scenario}
\vspace{-0.4 cm}
\end{figure}

\vspace{-0.1cm}
\section{System Model}\label{sec:system_model}

\subsection{Multi-View Wireless Sensing Scenario} \label{subsec:scenario}
{
We consider an uplink wireless sensing scenario as shown in Fig. \ref{fig:scenario}.
In this scenario, $B$ BSs are deployed, each equipped with a uniform linear array (ULA) of $N_r$ antennas, 
while $U$ single-antenna UEs are actively transmitting. 
Different from radar-centric systems, we realize the sensing function by reusing uplink channel estimation results   
and employ orthogonal frequency division multiplexing (OFDM) for signal modulation, which is designed to be compatible with communication systems. 
Each UE transmits pilot signals over $N_c$ subcarriers, which are scattered by the target within the RoI and subsequently received by all BSs.  
}Based on these pilots, each BS estimates the uplink CSI between itself and the corresponding UE. 
We define a transceiver pair as a single view. 
All BSs upload the CSI between each BS and all UEs to a central server, 
where all single-view CSI is combined to form the multi-view CSI for the subsequent sensing task.  

It is assumed that different UEs transmit pilots in separate time slots to avoid inter-user interference, 
and the positions of all BSs and UEs are known. 
{\color{black}The line-of-sight (LOS) path between the UE and BS does not contain environmental information, 
but it can be obtained based on the positions of the UE and BS, and then removed from the channel response.
}For simplicity, we will omit this part in the following analysis.

\vspace{-0.2cm}
\subsection{Signal Propagation and Channel Model} \label{subsec:channel_model}
{
We model the sensing channel based on rigorous EM scattering principles. 
}For convenience of presentation, we consider the two-dimensional (2-D) case with transverse-magnetic (TM) wave, 
where the electric field is oriented along the $\hat{z}$-axis and can be simplified as a scalar. 
We first analyze the system at the $n$-th subcarrier frequency $f_n$ and then extend to the multi-frequency case. 
Let $s_{u,n}$ denote the pilot signal transmitted by the $u$-th UE, and let $\mathcal{R}$ denote the RoI. 
For each position $\boldsymbol{r} \in \mathcal{R}$, the incident field generated by the $u$-th UE located at $\boldsymbol{p}^{\mathrm{UE}}_u$ is 
\begin{equation} \label{eq:continue_incident_field}
    E^{\mathrm{i}}_{u,n}(\boldsymbol{r}) = j k_n \eta \int g_{n}(\boldsymbol{r}, \boldsymbol{p}) \cdot \delta(\boldsymbol{p} - \boldsymbol{p}^{\mathrm{UE}}_u) s_{u,n} \; d\boldsymbol{p},
\vspace{-0.1cm}
\end{equation}
where $g_{n}(\boldsymbol{r},\boldsymbol{r}^{\prime}) = \frac{j}{4} H^{(1)}_{0}(k_n \|\boldsymbol{r} - \boldsymbol{r}^{\prime}\|)$ is the Green’s function in the 2-D case and $H_0^{(1)}(\cdot)$ denotes the zeroth order of Hankel function of the first kind.  
$k_n = 2\pi f_n/c$ is the wave number and $\eta = \sqrt{\mu_0/\varepsilon_0}$ is the characteristic impedance of free space \cite{kongElectromagneticWaveTheory2000}. 
The sensing targets are scattering media with specific shapes and EM characteristics. 
These properties are manifested as spatial distributions of relative permittivity and conductivity within the RoI, 
which can be compactly characterized using the contrast defined as 
\vspace{-0.1cm}
\begin{equation} \label{eq:continue_contrast}
    \chi_n(\boldsymbol{r}) = (\varepsilon_r(\boldsymbol{r}) - 1) + j\cdot \frac{\sigma(\boldsymbol{r})}{2\pi f_n \varepsilon_0}, \quad \text{for } \boldsymbol{r}\in\mathcal{R},
\vspace{-0.1cm}
\end{equation}
where $\varepsilon_r(\boldsymbol{r})$ and $\sigma(\boldsymbol{r})$ are the relative permittivity and conductivity at position $\boldsymbol{r}$, respectively. 
The wave-target interaction satisfies the following Lippmann-Schwinger equation \cite{chen2018computational}, 
\begin{equation} \label{eq:continue_state_eq}
    E^{\mathrm{t}}_{u,n}(\boldsymbol{r}) = E^{\mathrm{i}}_{u,n}(\boldsymbol{r}) + k_{n}^2 \int_{\mathcal{R}} g_{n}(\boldsymbol{r},\boldsymbol{r}^{\prime}) \chi_{n}(\boldsymbol{r}^{\prime}) E^{\mathrm{t}}_{u,n}(\boldsymbol{r}^{\prime}) d\boldsymbol{r}^{\prime}, 
\end{equation}
where $E^{\mathrm{t}}_{u,n}(\boldsymbol{r})$ is the total electric field at position $\boldsymbol{r}$. 
The scattered signals received by the BSs originate from the re-radiation of induced currents in the target. 
For the $r$-th antenna of the $b$-th BS, the received scattered field is
\begin{equation} \label{eq:continue_data_eq}
    E^{\mathrm{s}}_{b,r,u,n}(\boldsymbol{p}^{\mathrm{BS, RX}}_{b,r}) = k_n^2 \int_{\mathcal{R}} g_{n}(\boldsymbol{p}^{\mathrm{BS, RX}}_{b,r}, \boldsymbol{r}^{\prime}) \chi_{n}(\boldsymbol{r}^{\prime}) E^{\mathrm{t}}_{u,n}(\boldsymbol{r}^{\prime}) d\boldsymbol{r}^{\prime}, 
\end{equation}
where $\boldsymbol{p}^{\mathrm{BS, RX}}_{b,r}$ denotes the position of the corresponding receiving antenna. 

By uniformly discretizing the RoI into $D$ pixels, we can get its pixelized representation $\boldsymbol{x} = \left[\boldsymbol{\varepsilon}_r; \boldsymbol{\sigma}\right]$, 
where $\boldsymbol{\varepsilon}_r, \boldsymbol{\sigma}\in \mathbb{R}^{D\times 1}$ are the relative permittivity and conductivity at each pixel position. 
$\boldsymbol{x}$ is a dual-channel image that describes the target information in the RoI, and the discretized contrast image $\boldsymbol{\chi}\in\mathbb{C}^{D\times 1}$ can be obtained according to (\ref{eq:continue_contrast}).
By applying the method of moments (MoM) based on (\ref{eq:continue_incident_field})-(\ref{eq:continue_state_eq}), we can derive the discretized total electric field induced by the $u$-th UE as 
\begin{equation} \label{eq:discrete_total_field}
    \mathbf{e}^{\mathrm{t}}_{u,n} = \left[\mathbf{I} - \mathbf{G}_{n} \operatorname{diag}(\boldsymbol{\chi}_n)\right]^{-1} \mathbf{h}^{\mathrm{U-R}}_{u,n} s_{u,n}, 
\end{equation}
where $\mathbf{h}^{\mathrm{U-R}}_{u,n}\in\mathbb{C}^{D\times 1}$ denotes the spatial channel from the UE to RoI pixels, 
and $\mathbf{G}_{n}\in\mathbb{C}^{D\times D}$ is the discretized Green's function matrix for inter-pixel interactions. 
According to (\ref{eq:continue_data_eq}), the pilot signal received by the $b$-th BS on $N_r$ antennas is 
\begin{align}
    \mathbf{y}_{b,u,n} &= \mathbf{H}^{\mathrm{R-B}}_{b,n} \operatorname{diag}(\boldsymbol{\chi}_{n}) \mathbf{e}^{\mathrm{t}}_{u,n} + \mathbf{n} \\
    &= \mathbf{h}_{b,u,n} s_{u,n} + \mathbf{n}, \label{eq:y_hs}
\end{align}
where $\mathbf{H}^{\mathrm{R-B}}_{b,n}\in\mathbb{C}^{N_r\times D}$ denotes the spatial channel from RoI pixels to the BS, 
and $\mathbf{n}$ is additive white Gaussian noise. 
The specific forms of $\mathbf{G}_n$ can be derived from Maxwell's equations and the MoM, 
which are
\begin{equation}
\mathbf{G}_{n}[m, m^{\prime}] = \begin{cases}
    \begin{aligned}
        &\dfrac{jk_{n}\pi a}{2} J_{1}(k_{n}a) \\ 
        &\quad \times H_{0}^{(1)}\left(k_{n}\|\boldsymbol{r}_{m}-\boldsymbol{r}_{m^{\prime}}\|\right), \; \text{if } m \neq m^{\prime}, \\
        &\dfrac{jk_{n}\pi a}{2} H_{1}^{(1)}(k_{n}a)-1, \; \text{if } m = m^{\prime}, 
    \end{aligned}
\end{cases}
\end{equation}
where $\boldsymbol{r}_m$ denotes the position of the $m$-th RoI pixel. 
Each RoI pixel is approximated as a cicle with the same area, which has an equivalent radius $a$. 
$H_1^{(1)}(\cdot)$ and $J_1(\cdot)$ denote the first order of Hankel function of the first kind and the Bessel function of the first order, respectively \cite{chen2018computational}. 

According to (\ref{eq:discrete_total_field})-(\ref{eq:y_hs}), the spatial channel between the $u$-th UE and $b$-th BS is given by
\begin{align} \label{eq:SV_spatial_channel}
    \mathbf{h}_{b,u,n} = \mathbf{H}^{\mathrm{R-B}}_{b,n} \mathbf{X}_{n} \mathbf{h}^{\mathrm{U-R}}_{u,n} 
    = \left(({\mathbf{h}}^{\mathrm{U-R}}_{u,n})^\mathsf{T} \otimes {\mathbf{H}}^{\mathrm{R-B}}_{b,n}\right)\mathrm{vec}(\mathbf{X}_n). 
\vspace{-0.3cm}
\end{align}
where $\mathbf{X}_{n} = \operatorname{diag}(\boldsymbol{\chi}_{n}) \left[\mathbf{I} - \mathbf{G}_{n} \operatorname{diag}(\boldsymbol{\chi}_n)\right]^{-1}$ is the channel response induced by the target. 
By stacking $N_c$ subcarriers, we can obtain the vectorized form of spatial-frequency channel under a single view, 
\begin{align} \label{eq:single-view_H_vec}
    \mathbf{h}_{b,u} = \left[\mathbf{h}_{b,u,1}; \cdots; \mathbf{h}_{b,u,N_c}\right] 
    = \left((\breve{\mathbf{H}}^{\mathrm{U-R}}_{u})^{\mathsf{T}} \ast \breve{\mathbf{H}}^{\mathrm{R-B}}_{b}\right) \operatorname{vec}(\mathbf{X}).    
\end{align}
It consists of the following three components, 
\begin{align}
    \breve{\mathbf{H}}^{\mathrm{U-R}}_{u} &= \operatorname{diag}\left({\mathbf{h}}^{\mathrm{U-R}}_{u,1}, \cdots, {\mathbf{h}}^{\mathrm{U-R}}_{u,N_c}\right) = f_{\mathbf{H}_{\mathrm{U-R}}}(\boldsymbol{p}^{\mathrm{UE}}_u), \\
    \breve{\mathbf{H}}^{\mathrm{R-B}}_{b} &= \operatorname{diag}\left(\mathbf{H}^{\mathrm{R-B}}_{b,1}, \cdots, \mathbf{H}^{\mathrm{R-B}}_{b,N_c}\right) = f_{\mathbf{H}_{\mathrm{R-B}}}(\boldsymbol{p}^{\mathrm{BS}}_b), \\
    \mathbf{X} &= \left[\mathbf{X}_1, \cdots, \mathbf{X}_{N_c}\right] = f_{\mathbf{X}}(\boldsymbol{x}), 
\end{align}
which are related to the UE position $\boldsymbol{p}^{\mathrm{UE}}_u$, the BS position $\boldsymbol{p}^{\mathrm{BS}}_b$, and the target property $\boldsymbol{x}$, respectively.
The CSI matrix form of (\ref{eq:single-view_H_vec}) can be expressed as 
\begin{equation} \label{eq:single-view_H}
    \mathbf{H}_{b,u} = \operatorname{reshape}\left(\mathbf{h}_{b,u}, (N_r, N_c)\right) 
    = f_{\mathbf{H}}(\boldsymbol{x}, \boldsymbol{p}^{\mathrm{BS}}_b, \boldsymbol{p}^{\mathrm{UE}}_u). 
\end{equation}
We integrate the CSI from multiple views and the corresponding BS and UE positions into a unified set, 
\begin{equation} \label{eq:H_set}
\vspace{-0.1cm}
    \mathcal{H} = \left\{(\mathbf{H}_{b,u}, \boldsymbol{p}^{\mathrm{BS}}_b, \boldsymbol{p}^{\mathrm{UE}}_u) | b = 1, \cdots, B; u = 1, \cdots, U\right\}, 
\end{equation}
which forms the multi-view channel data.

\section{Conditional Generative learning for Multi-View Wireless Sensing} \label{sec:MV_gen}

\begin{figure*}[!t]
    \centering
    \includegraphics[width=0.85\linewidth]{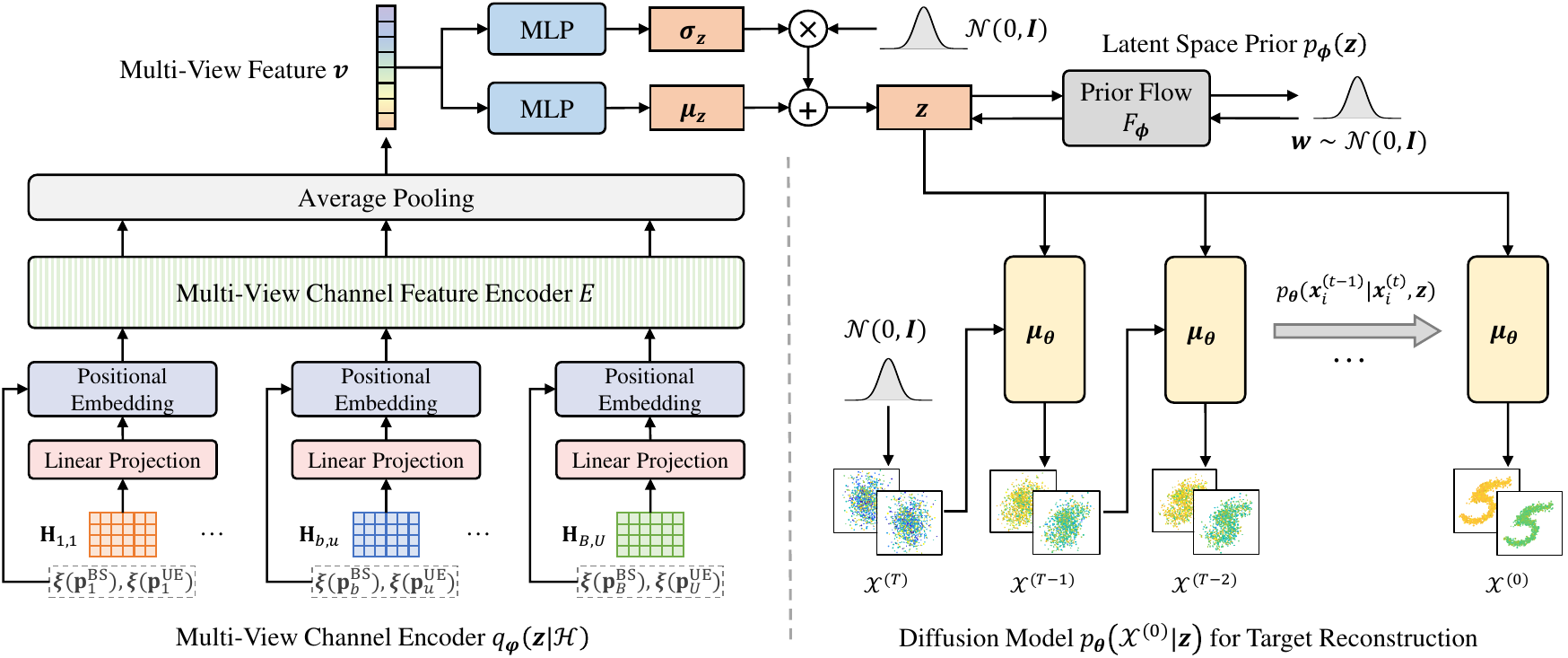}
    \vspace{-0.1cm}
    \caption{The proposed Gen-MV sensing framework,  
    which consists of a multi-view channel encoder $q_{\boldsymbol{\varphi}}(\boldsymbol{z}|\mathcal{H})$, 
    a latent space prior $p_{\phi}(\boldsymbol{z})$, and a conditional generative model $p_{\boldsymbol{\theta}}(\boldsymbol{x}|\boldsymbol{z})$. 
    Here, we employ a flow-based model $F_{\boldsymbol{\phi}}$ to parameterize the prior $p_{\boldsymbol{\phi}}(\boldsymbol{z})$, 
    and implement a conditional point cloud diffusion model $p_{\boldsymbol{\theta}}(\mathcal{X}^{(0)}|\boldsymbol{z})$ for target reconstruction, 
    where $\mathcal{X}^{(0)}$ is the point cloud representation of the target properties $\boldsymbol{x}$.}
    \vspace{-0.3cm}
    \label{fig:framework}
\end{figure*}

According to (\ref{eq:single-view_H}), the multi-view channels $\{\mathbf{H}_{b,u}\}$ share the same target property $\boldsymbol{x}$, 
which can be interpreted as projections of the identical target features from different BS/UE views. 
Therefore, joint processing of multi-view channel data can enhance target reconstruction. 
The objective of multi-view target reconstruction is to find the inverse mapping of (\ref{eq:single-view_H}), 
formulated as 
\begin{equation} \label{eq:inverse_mapping_pixel}
    \hat{\boldsymbol{x}} = g(\mathcal{H}), 
\end{equation}
which aims to recover the target properties from multi-view CSI and corresponding BS/UE positions. 
The inverse mapping $g$ can be implemented as an iterative algorithm or a classical AI model with deterministic input and output.  
Alternatively, it can also be designed based on powerful and flexible generative models to facilitate the learning of intrinsic physical properties from the data distribution.

In this section, we first introduce the proposed Gen-MV sensing framework, followed by the detailed design of the multi-view channel encoder and the target reconstruction scheme based on the diffusion model.

\vspace{-0.2cm}
\subsection{Gen-MV Sensing Framework} \label{subsec:gen_framework}
From the perspective of generative learning, we aim to model the conditional probability $p(\boldsymbol{x}|\mathcal{H})$ based on measurement data.  
It enables us to perform the reconstruction process in (\ref{eq:inverse_mapping_pixel}) through maximum a posteriori (MAP) or minimum mean square error (MMSE) estimation.
To achieve this, we can train a probabilistic model $p_{\boldsymbol{\theta}}(\boldsymbol{x}|\mathcal{H})$ with parameter $\boldsymbol{\theta}$ to approximate the true distribution $p(\boldsymbol{x}|\mathcal{H})$, 
where the training objective is to maximize the log-likelihood $\log p_{\boldsymbol{\theta}}(\boldsymbol{x}|\mathcal{H})$. 

However, maximizing the posterior distribution is generally intractable. 
According to the conditional variational autoencoder (CVAE) proposed in \cite{sohn2015CVAE}, we introduce a latent variable $\boldsymbol{z}$ and a learnable distribution $q_{\boldsymbol{\varphi}}(\boldsymbol{z}|\mathcal{H},\boldsymbol{x})$ to construct the evidence lower bound (ELBO) of $\log p_{\boldsymbol{\theta}}(\boldsymbol{x}|\mathcal{H})$, which is 
\begin{align}
    &\log{p_{\boldsymbol{\theta}}(\boldsymbol{x}|\mathcal{H})} 
    = \log\left(\mathbb{E}_{q_{\boldsymbol{\varphi}}(\boldsymbol{z}|\mathcal{H}, \boldsymbol{x})} \left\{\frac{p_{\boldsymbol{\theta}}(\boldsymbol{x}, \boldsymbol{z} | \mathcal{H})}{q_{\boldsymbol{\varphi}}(\boldsymbol{z}|\mathcal{H}, \boldsymbol{x})}\right\}\right) \nonumber \\
    &\stackrel{(a)}{\geq} \mathbb{E}_{q_{\boldsymbol{\varphi}}(\boldsymbol{z}|\mathcal{H}, \boldsymbol{x})} \log{\frac{p_{\boldsymbol{\theta}}(\boldsymbol{x}, \boldsymbol{z} | \mathcal{H})}{q_{\boldsymbol{\varphi}}(\boldsymbol{z}|\mathcal{H}, \boldsymbol{x})}}  \nonumber \\
    &= \mathbb{E}_{q_{\boldsymbol{\varphi}}(\boldsymbol{z}|\mathcal{H}, \boldsymbol{x})}\left[\log p_{\boldsymbol{\theta}}(\boldsymbol{x}|\boldsymbol{z},\mathcal{H}) - \log\frac{q_{\boldsymbol{\varphi}}(\boldsymbol{z}|\mathcal{H}, \boldsymbol{x})}{p_{\boldsymbol{\theta}}(\boldsymbol{z}|\mathcal{H})} \right] \nonumber \\
    &\triangleq \mathrm{ELBO}(\boldsymbol{x}, \mathcal{H}), 
\end{align}
where $(a)$ is due to Jensen's inequality. 
Therefore, we can implicitly optimize the log-likelihood $\log p_{\boldsymbol{\theta}}(\boldsymbol{x}|\mathcal{H})$ by minimizing the following loss function:  
\begin{equation} \label{eq:L_ELBO_original}
\begin{split}
    {L}_{\mathrm{ELBO}}(\boldsymbol{x}, \mathcal{H}) 
    &\triangleq \mathbb{E}_{q_{\boldsymbol{\varphi}}(\boldsymbol{z}|\mathcal{H}, \boldsymbol{x})}\left[-\log p_{\boldsymbol{\theta}}(\boldsymbol{x}|\boldsymbol{z},\mathcal{H})\right] \\
    &\quad + D_{\mathrm{KL}}\left(q_{\boldsymbol{\varphi}}(\boldsymbol{z}|\mathcal{H},\boldsymbol{x}) \parallel p_{\boldsymbol{\theta}}(\boldsymbol{z}|\mathcal{H})\right). 
\end{split}
\end{equation}

Although the above scheme is feasible for training a generative sensing model, 
we observe that it suffers from slow convergence and underperforming results in practice. 
{\color{black}
A potential reason for this issue is the model discrepancy between the training and inference stages in terms of architecture and task objectives. 
Specifically, in the training stage, since $\boldsymbol{x}$ is provided as a condition in $q_{\boldsymbol{\varphi}}(\boldsymbol{z}|\mathcal{H}, \boldsymbol{x})$, 
the first term of (\ref{eq:L_ELBO_original}) can reduce to compressing and reconstructing $\boldsymbol{x}$ using a variational autoencoder (VAE). 
This process is significantly easier than the inference task of reconstructing $\boldsymbol{x}$ from $\mathcal{H}$ based on $p_{\boldsymbol{\theta}}(\boldsymbol{z}|\mathcal{H})$ and $p_{\boldsymbol{\theta}}(\boldsymbol{x}|\boldsymbol{z},\mathcal{H})$. 
Although the KL divergence term in (\ref{eq:L_ELBO_original}) encourages alignment between $q_{\boldsymbol{\varphi}}(\boldsymbol{z}|\mathcal{H}, \boldsymbol{x})$ and $p_{\boldsymbol{\theta}}(\boldsymbol{z}|\mathcal{H})$, 
this inconsistency still negatively impacts the model's performance.
Moreover, the conditional prior model $q_{\boldsymbol{\varphi}}(\boldsymbol{z}|\mathcal{H}, \boldsymbol{x})$ serves only as a reference distribution during training and does not participate in the inference process, which introduces unnecessary redundancy to the model.
A discussion of similar potential issues in CVAE can also be found in \cite{sohn2015CVAE}.
}Our experiments indicate that relaxing (\ref{eq:L_ELBO_original}) to the following training objective leads to a more concise and efficient model: 
\vspace{-0.05cm}
\begin{equation} \label{eq:L_ELBO_relaxed}
\begin{split}
    \tilde{{L}}_{\mathrm{ELBO}}(\boldsymbol{x}, \mathcal{H}) 
    &\triangleq \mathbb{E}_{q_{\boldsymbol{\varphi}}(\boldsymbol{z}|\mathcal{H})}\left[-\log p_{\boldsymbol{\theta}}(\boldsymbol{x}|\boldsymbol{z})\right] \\ 
    &\quad + D_{\mathrm{KL}}\left(q_{\boldsymbol{\varphi}}(\boldsymbol{z}|\mathcal{H}) \parallel p_{\boldsymbol{\phi}}(\boldsymbol{z})\right). 
\end{split}
\vspace{-0.05cm}
\end{equation}
According to (\ref{eq:L_ELBO_relaxed}), we decompose the Gen-MV sensing process into the following two steps: 
\begin{itemize}
    \item \textbf{Step 1}: A multi-view channel encoder $q_{\boldsymbol{\varphi}}(\boldsymbol{z}|\mathcal{H})$ extracts the target latent code $\boldsymbol{z}$ from the multi-view channel data $\mathcal{H}$. 
    Inspired by VAE, we sample from a conditional distribution instead of using deterministic encoding and introduce a KL divergence term to align all posterior distributions with a common prior $p_{\boldsymbol{\phi}}(\boldsymbol{z})$, 
    ensuring a smooth and continuous latent space. 
    \item \textbf{Step 2}: A conditional generative model $p_{\boldsymbol{\theta}}(\boldsymbol{x}|\boldsymbol{z})$ generates the target representation from the latent code $\boldsymbol{z}$, thereby achieving the sensing of target properties. 
\end{itemize}
The former focuses on effectively extracting target features from multi-view channels, 
while the latter can be implemented by integrating conditioning mechanisms into generative \mbox{models}. 
This approach decouples the design of the channel encoder and the target reconstruction process, 
providing a flexible Gen-MV sensing framework. 
In Section \ref{sec:numerical_results}, we will experimentally demonstrate that the proposed simplified conditional generation framework outperforms the standard CVAE in the Gen-MV sensing task.

The schematic diagram of the overall framework is illustrated in Fig. \ref{fig:framework}. 
In the following subsections, we propose specific designs for the channel encoder and introduce a generative sensing scheme based on a diffusion model. 

\vspace{-0.2cm}

\subsection{Multi-View Channel Encoder} \label{subsec:MV_encoder}

\begin{figure*}[!htbp]
    \centering
    \hfill
    \begin{subfigure}[t]{0.185\linewidth}
        \includegraphics[width=\linewidth]{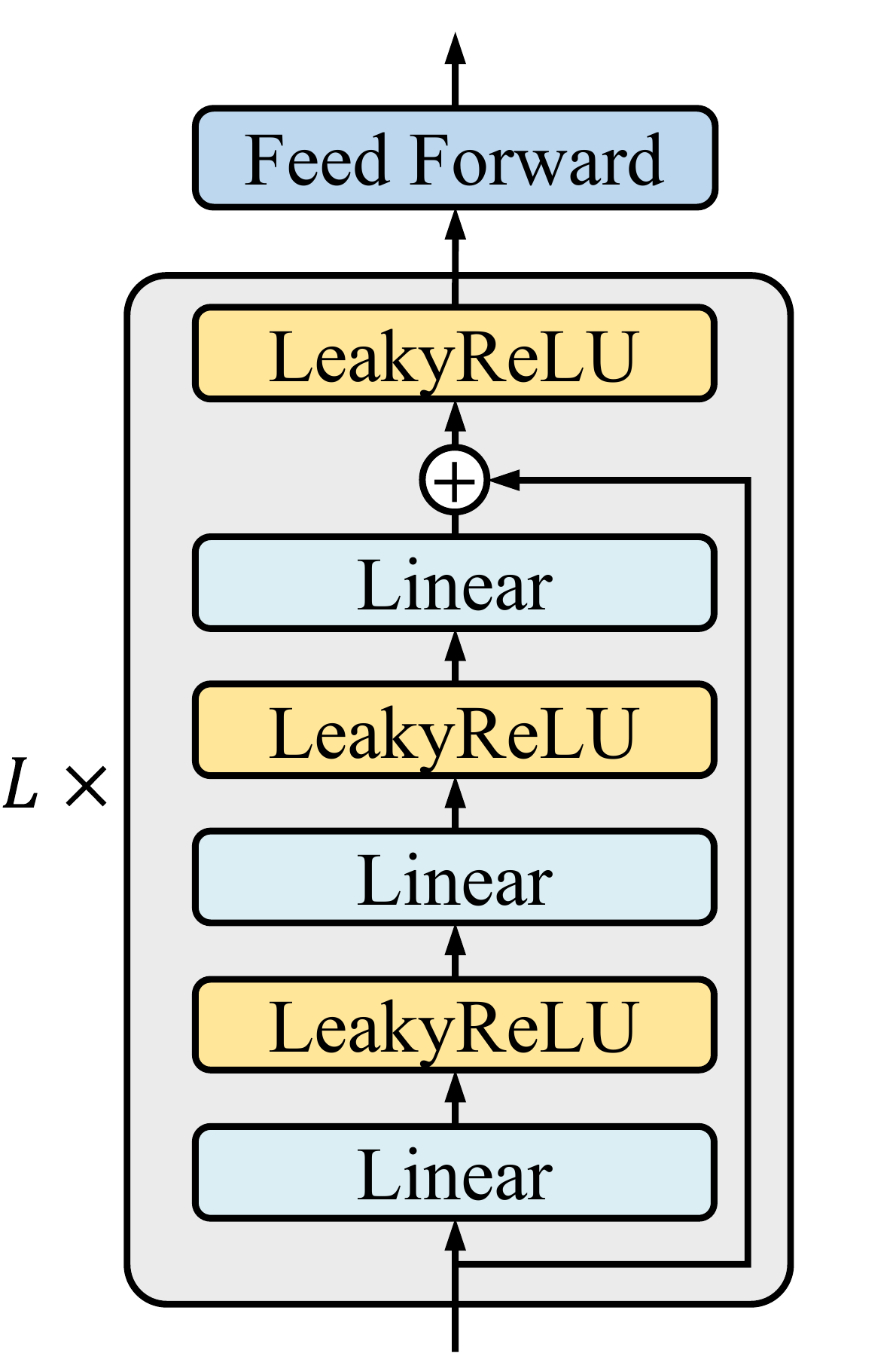}
        \caption{VS-MLP.}
        \label{fig:VS-MLP}
    \end{subfigure}
    \hfill
    \begin{subfigure}[t]{0.175\linewidth}
        \includegraphics[width=\linewidth]{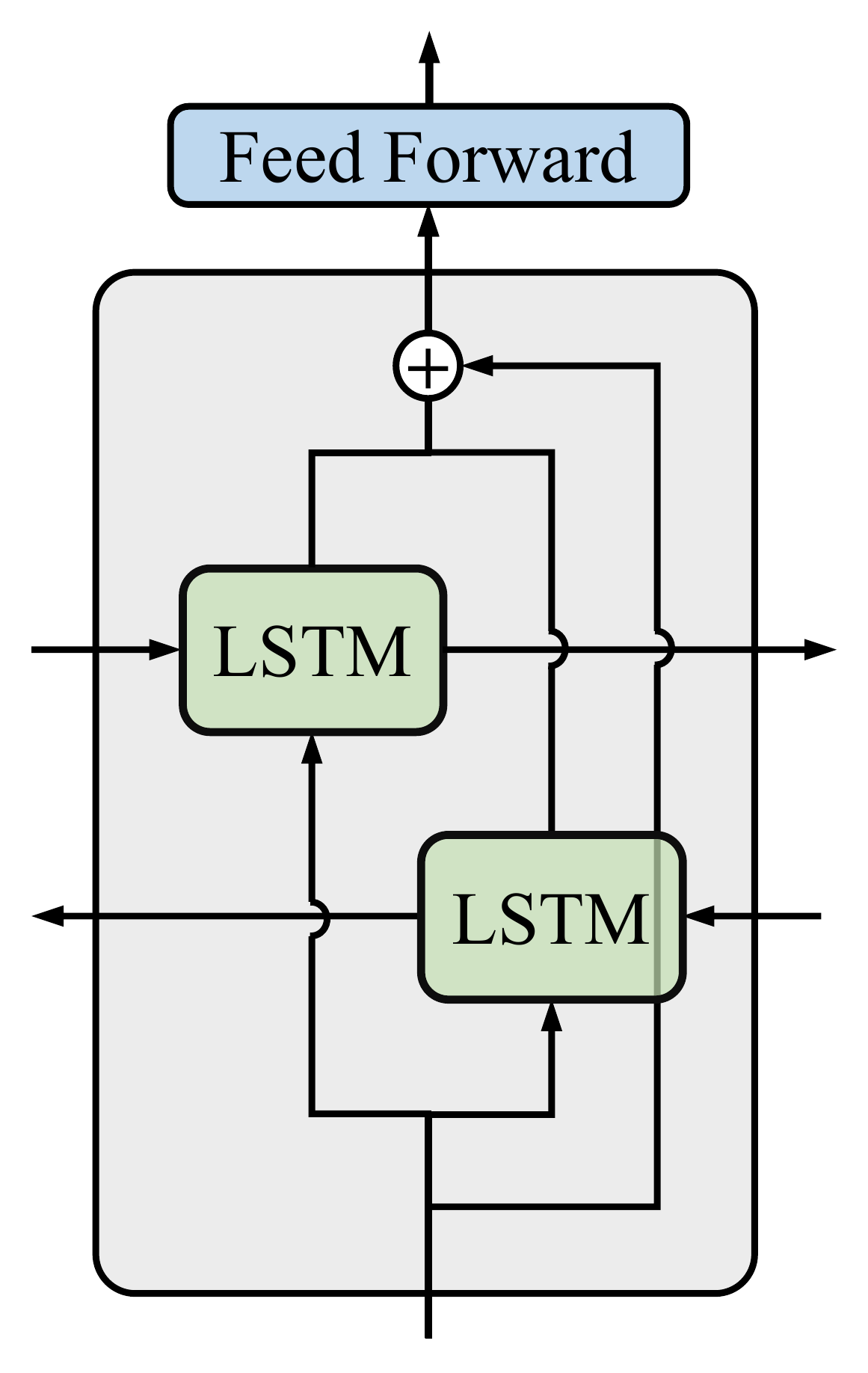}
        \caption{MV-BiLSTM.}
        \label{fig:MV-BiLSTM}
    \end{subfigure}
    \hfill
    \begin{subfigure}[t]{0.19\linewidth}
        \includegraphics[width=\linewidth]{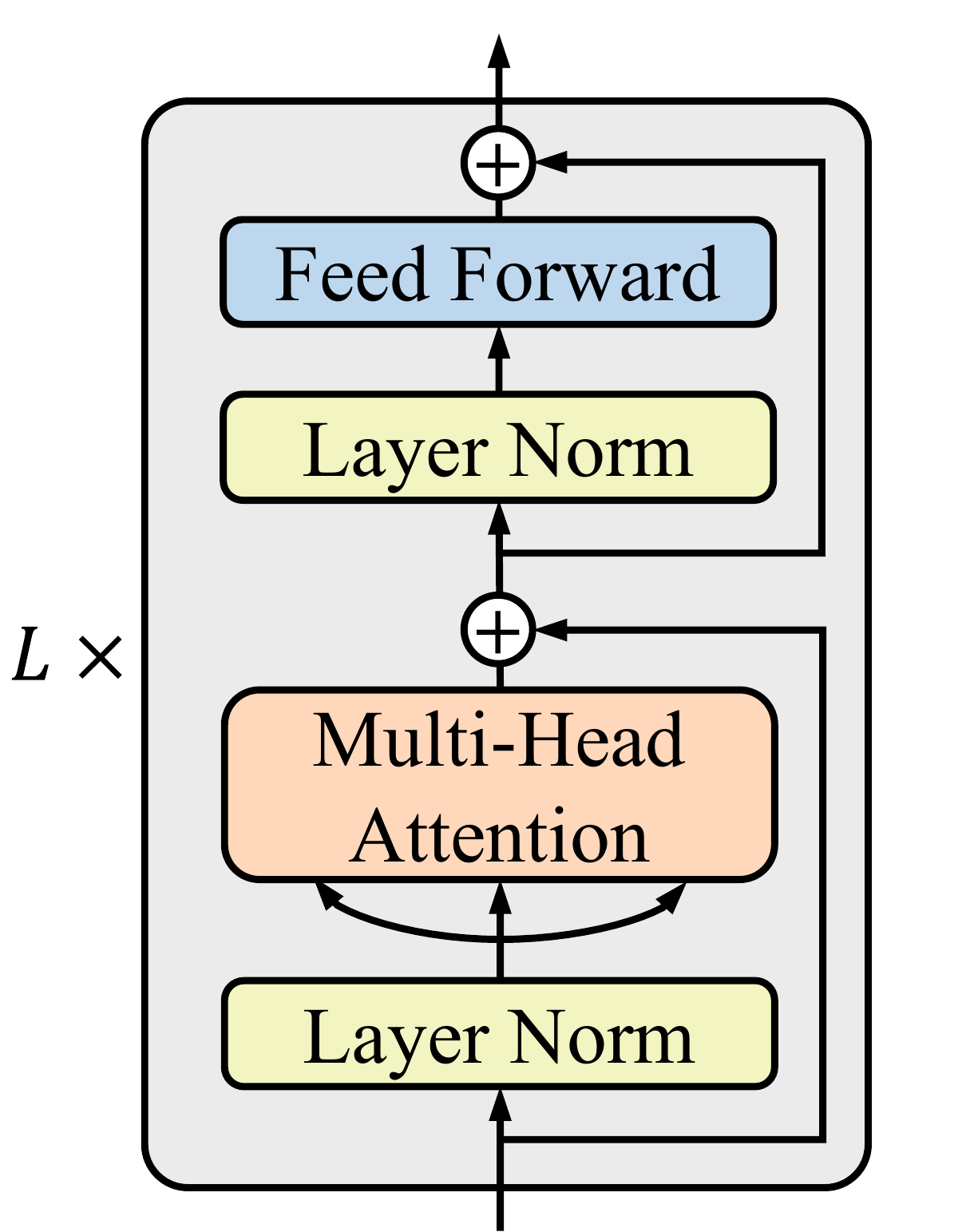}
        \caption{MVT.}
        \label{fig:MVT}
    \end{subfigure}
    \hfill
    \begin{subfigure}[t]{0.21\linewidth}
        \includegraphics[width=\linewidth]{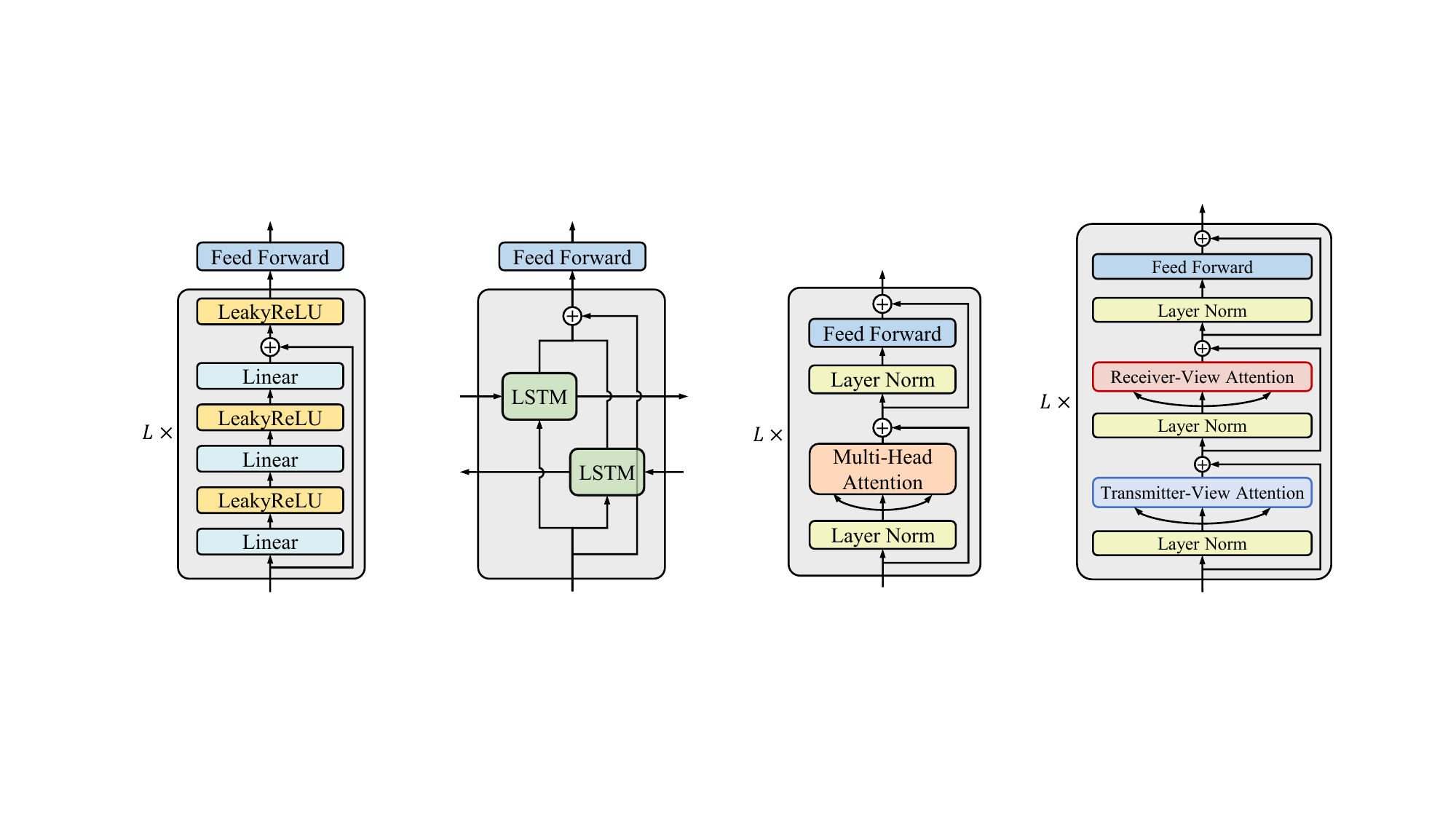}
        \caption{\color{black}IVT.}
        \label{fig:IVT}
    \end{subfigure}
    \hspace{0.6cm}
    \caption{\color{black}Four structural designs for the multi-view channel feature encoder $E$. 
    All feed-forward networks are two-layer MLPs with a hidden dimension of $2d_{\mathrm{v}}$. }
    \vspace{-0.3cm}
    \label{fig:4_encoders}
\end{figure*}

{ 
As shown in (\ref{eq:H_set}), the channel from each view needs to be combined with the BS and UE positions to fully characterize the single-view observation, which are scenario-dependent configurations. 
}For each single-view CSI, we flatten it and pass it through a fully-connected layer with parameters $\boldsymbol{W}_{\mathrm{h}}$ and $\boldsymbol{b}_{\mathrm{h}}$ to project it into a channel vector with dimension $d_{\mathrm{v}}$, 
\begin{equation} \label{eq:SV_channel_enc}
    \boldsymbol{h}_{b,u} = \boldsymbol{W}_{\mathrm{h}} \mathrm{vec}(\mathbf{H}_{b,u}) + \boldsymbol{b}_{\mathrm{h}}, \quad \forall b, u. 
\end{equation}
For the corresponding BS and UE positions, we first employ the positional encoding function in \cite{mildenhall2020nerf} to map spatial coordinates into higher dimensional vectors to capture the high-frequency variations of CSI in the space, i.e.  
\begin{equation}
\begin{split}
    \boldsymbol{\xi}(\boldsymbol{p}) = [\boldsymbol{p}; &\sin(2^0\pi \boldsymbol{p}); \cos(2^0\pi \boldsymbol{p}); \cdots; \\
    &\sin(2^{d_{\mathrm{p}}-1}\pi \boldsymbol{p}); \cos(2^{d_{\mathrm{p}}-1}\pi \boldsymbol{p})], 
\end{split}
\end{equation}
where $\boldsymbol{p}$ is the spatial coordinate and $d_{\mathrm{p}}$ denotes the number of encoding frequencies. 
By concatenating the encoded BS and UE position vectors, we can get the view vector to denote the view position, formulated as
\vspace{-0.05cm}
\begin{equation}
\vspace{-0.05cm}
    \boldsymbol{\xi}_{b,u} = \boldsymbol{\xi}(\boldsymbol{p}^{\mathrm{BS}}_{b}) \oplus \boldsymbol{\xi}(\boldsymbol{p}^{\mathrm{UE}}_{u}). 
\end{equation}

Embedding the view position into the channel vector is crucial and fundamentally different from the positional embedding in natural language processing (NLP). 
In natural language, the meaning of a word is relatively independent of its position in the text. 
Additionally, word embeddings and their positional encodings are inherently discrete and can be regarded as transformed representations of one-hot vectors. 
Due to the approximate orthogonality of random high-dimensional vectors, NLP commonly employs additive positional embeddings, enabling the model to separate lexical and positional information in later stages.  
In contrast, the wireless channel is highly correlated with its view position, and both are physical quantities with continuous values. 
These factors challenge the adaptability of additive positional embedding in wireless sensing models.
According to (\ref{eq:single-view_H_vec}), the channel $\mathbf{h}_{b,u}$ is formed by the coupling of $((\breve{\mathbf{H}}^{\mathrm{U-R}}_{u})^{\mathsf{T}} \ast \breve{\mathbf{H}}^{\mathrm{R-B}}_{b})$ and $\mathbf{X}$ through a linear transformation, 
where the two terms respectively contain the information of the view position and the target. 
Inspired by this, we construct a linear transformation to decouple the channel vector from the view position, formulated as
\vspace{-0.05cm}
\begin{equation} \label{eq:pos_embed}
\vspace{-0.05cm}
    \boldsymbol{h}^{\mathrm{in}}_{b,u} = \boldsymbol{\Gamma}\left(\boldsymbol{\xi}_{b,u}\right) \cdot \boldsymbol{h}_{b,u} = \boldsymbol{\gamma}\left(\boldsymbol{\xi}_{b,u}\right) \odot \boldsymbol{h}_{b,u}, \quad \forall b, u, 
\end{equation}
where $\boldsymbol{\Gamma}$ is a learnable transformation matrix derived from the view vector $\boldsymbol{\xi}_{b,u}$ and is assumed to be diagonal for easier model learning, which can be simplified to a multiplicative positional embedding. 
$\boldsymbol{\gamma}(\cdot)$ is a fully-connected layer with input-output dimensions of $4\cdot(2d_{\mathrm{p}}+1) \rightarrow d_{\mathrm{v}}$, 
and $\{\boldsymbol{h}^{\mathrm{in}}_{b,u}\}$ are channel vectors after positional embedding. 

To achieve a comprehensive measurement of the scattered field for high-quality sensing, it is essential to fuse CSI from multiple views to obtain the target latent code $\boldsymbol{z}$. 
First, we extract features from the embedded channel vectors through a multi-view channel feature encoder $E$, which can be formulated generally as
\vspace{-0.05cm}
\begin{equation} \label{eq:multi-view_channel_feature_enc}
    \left\{\boldsymbol{h}^{\mathrm{out}}_{b,u} \, | \, \forall b,u\right\} = E \left(\{\boldsymbol{h}^{\mathrm{in}}_{b,u} \, | \, \forall b,u \}\right). 
\vspace{-0.05cm}
\end{equation}
Through average pooling, the channel features from all views are aggregated into a multi-view feature vector $\boldsymbol{v}$, 
\vspace{-0.05cm}
\begin{equation} \label{eq:v_avg_pool}
    \boldsymbol{v} = \frac{1}{B\cdot U}\sum_{b, u} \boldsymbol{h}^{\mathrm{out}}_{b,u}. 
\vspace{-0.05cm}
\end{equation} 
Finally, the posterior distribution of the target latent code $\boldsymbol{z}$ is modeled as  
\vspace{-0.05cm}
\begin{equation} \label{eq:z_v}
    \boldsymbol{z} \sim \mathcal{N}\left(\boldsymbol{z}; \boldsymbol{\mu}_{\mathrm{z}}(\boldsymbol{v}), \mathrm{diag}(\boldsymbol{\sigma}_{\mathrm{z}}(\boldsymbol{v}))\right), 
\vspace{-0.05cm}
\end{equation}
where $\boldsymbol{\mu}_{\mathrm{z}}(\cdot)$ and $\boldsymbol{\sigma}_{\mathrm{z}}(\cdot)$ are shallow multilayer perceptrons (MLPs) with input-output dimensions of $d_{\mathrm{v}}\rightarrow d_{\mathrm{z}}$, and the sampling of $\boldsymbol{z}$ is performed via the reparameterization trick. 

The complete multi-view channel encoder $q_{\boldsymbol{\varphi}}(\boldsymbol{z}|\mathcal{H})$ is constructed by cascading (\ref{eq:SV_channel_enc})-(\ref{eq:z_v}), as shown in the left part of Fig. \ref{fig:framework}.  
With the multi-view channel feature encoder $E$ serving as the core module, its design determines the effectiveness of multi-view fusion. 
Below, we implement the encoder $E$ using three classical neural networks as baselines, 
and further propose an interleaved correlation learning architecture specifically designed for multi-view channel data. 

\subsubsection{View-Shared MLP (VS-MLP)}
According to (\ref{eq:single-view_H}), all single-view channels share the identical propagation function $f_{\mathbf{H}}$ and target property $\boldsymbol{x}$. 
Leveraging this characteristic, a basic encoder $E$ processes each single-view channel in parallel through a shared feature extractor, 
with multi-view information fusion entirely handled by subsequent average pooling. 
For single-view feature extraction, we utilize a MLP shown in Fig. \ref{fig:VS-MLP}, with its parameters shared across all views. 

\subsubsection{Multi-View BiLSTM (MV-BiLSTM)}
Although VS-MLP is effective and parameter-efficient, average pooling is insufficient for fusing multi-view information. 
In \cite{chen2023deep}, the authors achieved high-precision localization by accumulating channel features across multiple frequencies using Long Short Term Memory (LSTM). 
Inspired by this, we process $\{\boldsymbol{h}^{\mathrm{in}}_{b,u}\}$ as a sequence of length $B\times U$ and design a Bidirectional LSTM (BiLSTM) in Fig. \ref{fig:MV-BiLSTM} as the encoder $E$ to integrate multi-view channel features. 
Compared to LSTM, BiLSTM introduces bidirectional feature accumulation, which enables each output feature to incorporate information from all views, enhancing compatibility with subsequent average pooling. 

\subsubsection{Multi-View Transformer (MVT)}
As shown in (\ref{eq:H_set}), the multi-view channel data $\mathcal{H}$ is an unordered set. 
However, MV-BiLSTM is a sequential model and cannot guarantee permutation invariance of the multi-view channel encoder. 
To address this, we designed the MVT, as illustrated in Fig. \ref{fig:MVT}. 
MVT treats $\{\boldsymbol{h}^{\mathrm{in}}_{b,u}\}$ as a set with $B\times U$ elements and computes pairwise correlations between channel vectors across all views through self-attention for feature extraction. 
It resolves the long-range dependency issue in LSTMs, and ensures permutation equivariance between input and output. 
With subsequent average pooling, the MVT-based multi-view channel encoder strictly satisfies permutation invariance. 

\subsubsection{Interleaved-View Transformer (IVT)}
The three classical architectures mentioned above can all be utilized for multi-view channel feature encoding, and we implement them as baseline models. 
However, they still flatten the multi-view channel data $\mathcal{H}$ into a one-dimensional sequence or set for processing, ignoring the inherent structure of multi-view CSI. 
Specifically, as shown in (\ref{eq:single-view_H_vec}), each single-view channel $\mathbf{H}_{b,u}$ comprises of three components: 
the subchannel $\breve{\mathbf{H}}^{\mathrm{U-R}}_{u}$ determined by the UE position, 
the subchannel $\breve{\mathbf{H}}^{\mathrm{R-B}}_{b}$ determined by the BS position, 
and the channel response $\mathbf{X}$ induced by the scattering targets. 
To reveal the intrinsic structure of multi-view channels, we can organize the channels across all views into a block matrix as illustrated below, 
\vspace{-0.1cm}
\begin{align}
    \mathbf{H}_{\mathrm{mv}} &= 
    \begin{bmatrix}
        \mathbf{H}_{1,1} & \cdots & \mathbf{H}_{1,U} \\
        \vdots  & \ddots & \vdots \\
        \mathbf{H}_{B,1} & \cdots & \mathbf{H}_{B,U}
    \end{bmatrix}\\
    &= \begin{bmatrix}
        f_{\mathbf{H}}(\boldsymbol{x}, \boldsymbol{p}^{\mathrm{BS}}_1, \boldsymbol{p}^{\mathrm{UE}}_1) & \cdots & f_{\mathbf{H}}(\boldsymbol{x}, \boldsymbol{p}^{\mathrm{BS}}_1, \boldsymbol{p}^{\mathrm{UE}}_U) \\
        \vdots & \ddots & \vdots \\
        f_{\mathbf{H}}(\boldsymbol{x}, \boldsymbol{p}^{\mathrm{BS}}_B, \boldsymbol{p}^{\mathrm{UE}}_1) & \cdots & f_{\mathbf{H}}(\boldsymbol{x}, \boldsymbol{p}^{\mathrm{BS}}_B, \boldsymbol{p}^{\mathrm{UE}}_U)
    \end{bmatrix}, 
\end{align}
where $f_{\mathbf{H}}$ is the single-view channel function defined in (\ref{eq:single-view_H}). 
It can be observed that each row block's $U$ channels reflect the variation of CSI as the UE position changes under a fixed BS view. 
According to (\ref{eq:single-view_H_vec}), this corresponds to multiple single-view channels generated by different $\breve{\mathbf{H}}^{\mathrm{U-R}}_{u}$ with a shared $\breve{\mathbf{H}}^{\mathrm{R-B}}_{b}$, and this property is consistent across all BSs. 
Similarly, each column block's $B$ channels reflect the CSI variation caused by changes in the BS position with a fixed UE view, 
corresponding to multiple single-view channels with different $\breve{\mathbf{H}}^{\mathrm{R-B}}_{b}$ with a shared $\breve{\mathbf{H}}^{\mathrm{U-R}}_{u}$, and it is a property consistent across all UEs. 

Inspired by the interleaved learning method in \cite{chen2024mixer}, 
and based on the structural properties of the multi-view channels as well as the coupling form of $\breve{\mathbf{H}}^{\mathrm{U-R}}_{u}$ and $\breve{\mathbf{H}}^{\mathrm{R-B}}_{b}$ shown in (\ref{eq:single-view_H_vec}), 
we propose a multi-view channel encoder named IVT, which alternately extracts feature correlations from the UE and BS views, with its model structure illustrated in Fig. \ref{fig:IVT}. 
In detail, each IVT layer operates through: 
\begin{itemize}
    \item Transmitter-View Attention (TVA): Computes channel feature correlations between different UE views, with this operation being shared across all BS views. 
    \item Receiver-View Attention (RVA): Computes channel feature correlations between different BS views, with this operation being shared across all UE views. 
    \item Feed-Forward Network (FFN): Performs nonlinear transformations to enable hierarchical channel feature extraction through stacked model layers. 
\end{itemize}
The input $\{\boldsymbol{h}^{\mathrm{in}}_{b,u}\}$ can be structured as a $B\times U\times d_{\mathrm{v}}$ tensor $\boldsymbol{h}_{0}$. 
For the $\ell$-th IVT layer, the computation process is formulated as
\begin{align}
    &\boldsymbol{h}_{\ell} [b, :, :] = \boldsymbol{h}_{\ell-1} [b, :, :] + \mathrm{TVA}(\mathrm{LN}(\boldsymbol{h}_{\ell-1} [b, :, :])), \enspace \forall b, \\
    &\boldsymbol{h}_{\ell} [:, u, :] = \boldsymbol{h}_{\ell} [:, u, :] + \mathrm{RVA}(\mathrm{LN}(\boldsymbol{h}_{\ell} [:, u, :])), \enspace \forall u, \\
    &\boldsymbol{h}_{\ell} [b, u, :] = \boldsymbol{h}_{\ell} [b, u, :] + \mathrm{FFN}(\mathrm{LN}(\boldsymbol{h}_{\ell} [b, u, :])), \enspace \forall b,u, 
\end{align}
where $\ell=1,\cdots,L$ and $L$ is the number of IVT layers. 
$\mathrm{TVA}(\cdot)$ and $\mathrm{RVA}(\cdot)$ are self-attention blocks that extract correlations along the second and first dimensions of $\boldsymbol{h}_{(\cdot)}$ respectively, 
and $\mathrm{LN}(\cdot)$ denotes layer normalization. 

By alternately analyzing the channel feature correlations along the transmitter and receiver view dimensions, 
IVT incorporate the prior knowledge of multi-view channel structure into its model architecture. 
This approach better aligns the encoder's computation with the characteristics of the channel data, 
enabling to extract the embedded target information $\boldsymbol{x}$ from all view's channel $\mathcal{H}$ more efficiently. 

Overall, the aforementioned four encoders process the multi-view channel data $\mathcal{H}$ as four distinct data structures: 
VS-MLP treats them as independent observations, MV-BiLSTM processes them as sequential data, MVT organizes them as an unordered set, and IVT models them as a 2-D interleaved collection. 
In Section \ref{sec:numerical_results}, we will compare the effectiveness of these four model structures in extracting multi-view features. 

\subsection{Diffusion Model for Target Reconstruction} \label{subsec:diffusion}
Since the target latent code $\boldsymbol{z}$ extracted by the multi-view channel encoder contains the complete properties of the target, 
it is feasible to generate the corresponding target representation using appropriate conditional generative models. 
Different from the pixel-based representation $\boldsymbol{x}$ in forward modeling, in the reconstruction process, 
we represent the target as a \mbox{4-D} point cloud containing $M$ normalized shape-EM points, 
$\mathcal{X}^{(0)} = \{\boldsymbol{x}^{(0)}_i | i = 1,\cdots,M\}$. 
Each point is defined as 
\begin{equation} \label{eq:normalized_point}
    \boldsymbol{x}^{(0)}_i = \left[\frac{x_i-\mu_x}{v_{x}}, \frac{y_i-\mu_y}{v_{y}}, \frac{\varepsilon_i-\mu_{\varepsilon}}{v_{\varepsilon}}, \frac{\sigma_i-\mu_{\sigma}}{v_{\sigma}}\right]^{\mathsf{T}}, 
\end{equation}
where $x_i$, $y_i$, $\varepsilon_i$, $\sigma_i$ denote the $\hat{x}$-axis coordinate, $\hat{y}$-axis coordinate, relative permittivity and conductivity of the $i$-th point, respectively, 
while $\mu_{(\cdot)}$ and $v_{(\cdot)}$ respectively denote the mean and standard deviation of the corresponding dimensions. 

Compared to the pixel-based target representation $\boldsymbol{x}$ in Section \ref{subsec:channel_model}, 
the point cloud $\mathcal{X}^{(0)}$ offers the following advantages: 
\begin{itemize}
    \item \textbf{Low redundancy}: Due to the sparsity of target distribution in the RoI, 
    $\boldsymbol{x}$ contains numerous background pixels, whereas the point cloud $\mathcal{X}^{(0)}$ only needs to represent the target and omits background information, resulting in reduced redundancy. 
    \item \textbf{Probabilistic characteristic}: The point cloud $\mathcal{X}^{(0)}$ emphasizes the holistic distribution of points, 
    which can be modeled as $M$ independent samples from a point distribution $q(\boldsymbol{x}^{(0)}_i|\boldsymbol{z})$. 
    This intrinsic stochasticity makes the point cloud more suitable for probabilistic generative models in target reconstruction. 
\end{itemize}
Moreover, using distinct target representations for forward modeling and inverse reconstruction requires models to learn implicitly embedded physical knowledge from data. 
This strategy eliminates the dependence on explicit and precise forward modeling in the reconstruction process. 

In this subsection, we adopt a conditional point cloud diffusion model $p_{\boldsymbol{\theta}}(\mathcal{X}^{(0)}|\boldsymbol{z})$ proposed in \cite{luo2021diffusion} to generate the reconstructed target based on the target latent code $\boldsymbol{z}$, 
and derive the optimization procedure of the complete model based on the proposed Gen-MV sensing framework.

\subsubsection{Forward Diffusion Process}
The forward diffusion process gradually transforms the original point cloud distribution $q(\boldsymbol{x}^{(0)}_i)$ into a noise distribution $q(\boldsymbol{x}^{(T)}_i)$.
It can be modeled as a Markov chain, 
\vspace{-0.05cm}
\begin{equation} \label{eq:forward_Markov}
\vspace{-0.05cm}
    q(\boldsymbol{x}_i^{(1:T)}|\boldsymbol{x}_i^{(0)})=\prod_{t=1}^Tq(\boldsymbol{x}_i^{(t)}|\boldsymbol{x}_i^{(t-1)}), 
\end{equation}
and the transition probability is defined as 
\vspace{-0.05cm}
\begin{equation} \label{eq:forward_one_step}
\vspace{-0.05cm}
    q(\boldsymbol{x}^{(t)}_i|\boldsymbol{x}^{(t-1)}_i)=\mathcal{N}\left(\boldsymbol{x}^{(t)}_i;\sqrt{1-\beta_t}\boldsymbol{x}^{(t-1)}_i,\beta_t\mathbf{I}\right),  
\end{equation}
where $t=1,\cdots,T$, with $T$ being the maximum timestep in the diffusion process. 
The predefined noise variance schedule $\beta_1,\cdots,\beta_T$ controls the rate of the diffusion. 
Defining $\alpha_t\triangleq 1-\beta_t$, $\bar{\alpha}_t\triangleq\prod_{s=1}^t\alpha_s$, 
the sampling of $\boldsymbol{x}^{(t)}$ in the forward diffusion can derived from (\ref{eq:forward_Markov}) and (\ref{eq:forward_one_step}) as
\vspace{-0.05cm}
\begin{equation} \label{eq:forward_xt}
\vspace{-0.05cm}
    \boldsymbol{x}_i^{(t)}=\sqrt{\bar{\alpha}_t}\boldsymbol{x}_i^{(0)} + \sqrt{1-\bar{\alpha}_t} \boldsymbol{\epsilon}^{(t)}_i, \quad \boldsymbol{\epsilon}^{(t)}_i\sim \mathcal{N}\left(\boldsymbol{0}, \mathbf{I}\right). 
\end{equation}

\subsubsection{Reverse Diffusion Process} \label{subsubsec:reverse_diffusion}
The generation process is the reverse of the forward diffusion process, 
aiming to gradually generate a target point cloud from the noise distribution conditioned on the target latent code $\boldsymbol{z}$. 

According to Bayes' theorem, the reverse transition probability conditioned on $\boldsymbol{x}^{(0)}_i$ can be derived from (\ref{eq:forward_one_step}) and (\ref{eq:forward_xt}) as 
\vspace{-0.05cm}
\begin{equation} \label{eq:reverse_one_step_x0}
\vspace{-0.05cm}
    q(\boldsymbol{x}^{(t-1)}_{i} | \boldsymbol{x}^{(t)}_{i}, \boldsymbol{x}^{(0)}_{i}) = \mathcal{N}\left(\boldsymbol{x}^{(t-1)}_{i}; \tilde{\boldsymbol{\mu}}_t(\boldsymbol{x}^{(t)}_{i}, \boldsymbol{x}^{(0)}_{i}), \tilde{\beta}_t\mathbf{I}\right),
\end{equation}
with
\vspace{-0.05cm}
\begin{empheq}[left=\empheqlbrace]{align}
\vspace{-0.05cm}
    &\tilde{\boldsymbol{\mu}}_t(\boldsymbol{x}^{(t)}_{i}, \boldsymbol{x}^{(0)}_{i}) = \frac{1}{\sqrt{\alpha_t}} \Big( \boldsymbol{x}^{(t)}_{i} - \frac{1 - \alpha_t}{\sqrt{1 - \bar{\alpha}_t}} \boldsymbol{\epsilon}^{(t)}_{i} \Big),  \label{eq:reverse_x0_mu} \\
    &\tilde{\beta}_t = \frac{1-\bar{\alpha}_{t-1}}{1-\bar{\alpha}_{t}} \beta_t,  \label{eq:reverse_x0_sigma} 
\end{empheq}
where $\boldsymbol{\epsilon}^{(t)}_{i} = (\boldsymbol{x}_i^{(t)}-\sqrt{\bar{\alpha}_t}\boldsymbol{x}_i^{(0)}) / \sqrt{1-\bar{\alpha}_t}$.  
However, the $q(\boldsymbol{x}^{(t-1)}_{i} | \boldsymbol{x}^{(t)}_{i}, \boldsymbol{z})$ we needed is still intractable. 
Therefore, we need to construct a learnable Markov chain conditioned on $\boldsymbol{z}$ to implement the reverse diffusion process, 
where input points are sampled from the noise distribution $p(\boldsymbol{x}^{(T)}_i)$ to approximate $q(\boldsymbol{x}^{(T)}_i)$, i.e. 
\vspace{-0.05cm}
\begin{equation} \label{eq:reverse_Markov}
\vspace{-0.05cm}
    p_{{\boldsymbol{\theta}}}(\boldsymbol{x}^{(0:T)}_i|\boldsymbol{z})=p(\boldsymbol{x}^{(T)}_i)\prod_{t=1}^Tp_{{\boldsymbol{\theta}}}(\boldsymbol{x}^{(t-1)}_i|\boldsymbol{x}^{(t)}_i,\boldsymbol{z}),
\end{equation}
and the reverse transition probability is modeled as a learnable \mbox{Gaussian} distribution, 
\begin{equation} \label{eq:reverse_one_step}
    p_{{\boldsymbol{\theta}}}(\boldsymbol{x}^{(t-1)}_i|\boldsymbol{x}^{(t)}_i,\boldsymbol{z})=\mathcal{N}\left(\boldsymbol{x}^{(t-1)}_i;\boldsymbol{\mu}_{{\boldsymbol{\theta}}}(\boldsymbol{x}^{(t)}_i,t,\boldsymbol{z}),\tilde{\beta}_t\mathbf{I}\right). 
\end{equation}
Similar to $\tilde{\boldsymbol{\mu}}_t$ in (\ref{eq:reverse_x0_mu}), the predicted mean $\boldsymbol{\mu}_{{\boldsymbol{\theta}}}$ is defined as 
\begin{equation} \label{eq:mu_theta}
    \boldsymbol{\mu}_{{\boldsymbol{\theta}}}(\boldsymbol{x}^{(t)}_i, t, \boldsymbol{z}) = \frac{1}{\sqrt{\alpha_t}} \left( \boldsymbol{x}^{(t)}_i - \frac{1 - \alpha_t}{\sqrt{1 - \bar{\alpha}_t}} \boldsymbol{\epsilon}_{{\boldsymbol{\theta}}}(\boldsymbol{x}^{(t)}_i, t, \boldsymbol{z}) \right),  
\end{equation}
where $\boldsymbol{\epsilon}_{{\boldsymbol{\theta}}}$ is a learnable noise predictor with parameter $\boldsymbol{\theta}$. 
Following \cite{luo2021diffusion}, we construct $\boldsymbol{\epsilon}_{\boldsymbol{\theta}}$ using $L_{\mathrm{d}}$ cascaded ConcatSquash layers \cite{grathwohl2018ffjord}, with the $\ell$-th layer computed as:
\begin{equation}
\begin{split}
    &\boldsymbol{h}_{\mathrm{d}}^{(\ell)} = \mathrm{CS}(\boldsymbol{h}_{\mathrm{d}}^{(\ell-1)}, t, \boldsymbol{z}) \\ 
    &= (\boldsymbol{W}_{1}^{(\ell)} \boldsymbol{h}_{\mathrm{d}}^{(\ell-1)} + \boldsymbol{b}_1^{(\ell)}) \odot \sigma(\boldsymbol{W}_{2}^{(\ell)} \boldsymbol{c} + \boldsymbol{b}_2^{(\ell)}) + \boldsymbol{W}_{3}^{(\ell)} \boldsymbol{c},  
\end{split}
\end{equation}
where $\boldsymbol{h}_{\mathrm{d}}^{(\ell-1)}$ and $\boldsymbol{h}_{\mathrm{d}}^{(\ell)}$ are the input and output feature vectors of the $\ell$-th layer of the noise estimation network $\boldsymbol{\epsilon}_{\boldsymbol{\theta}}$. 
The conditional vector is defined as $\boldsymbol{c} = [\boldsymbol{\zeta}_{(t)}; \boldsymbol{z}]$, where $\boldsymbol{\zeta}_{(t)}$ is the $d_{\mathrm{t}}$-dimensional encoding of the timestep $t$, 
with its elements given by $\boldsymbol{\zeta}_{(t)}[2i] = \sin(t / 10000^{2i/d_t})$ and $\boldsymbol{\zeta}_{(t)}[2i + 1] = \cos(t / 10000^{2i/d_t})$. 
$\boldsymbol{W}_{1}^{(\ell)}$, $\boldsymbol{W}_{2}^{(\ell)}$, $\boldsymbol{W}_{3}^{(\ell)}$, $\boldsymbol{b}_1^{(\ell)}$, and $\boldsymbol{b}_2^{(\ell)}$ are learnable parameters, and $\sigma(\cdot)$ denotes the Sigmoid function. 
LeakyReLU is used as the activation function between two adjacent ConcatSquash layers. 
The dimensionalities of input $\boldsymbol{h}_{\mathrm{d}}^{(0)}$ and output $\boldsymbol{h}_{\mathrm{d}}^{(L_{\mathrm{d}})}$ of $\boldsymbol{\epsilon}_{\boldsymbol{\theta}}$ match that of the 4-D point cloud.

\subsubsection{\color{black}Optimization of the Diffusion Model} \label{subsubsec:optim_diffusion}
For a given target point cloud $\mathcal{X}^{(0)}$ and its latent code $\boldsymbol{z}$, the training objective of the point cloud diffusion model is to maximize the log-likelihood $\log{p_{{\boldsymbol{\theta}}}(\mathcal{X}^{(0)}|\boldsymbol{z})}$. 
By introducing the point clouds $\mathcal{X}^{(1:T)}$ in the diffusion process, we can construct the ELBO of $\log{p_{{\boldsymbol{\theta}}}(\mathcal{X}^{(0)}|\boldsymbol{z})}$ as 
\begin{align} \label{eq:neg_log_likelihood}
    &\log{p_{{\boldsymbol{\theta}}}(\mathcal{X}^{(0)}|\boldsymbol{z})} = \log\left[\mathbb{E}_{q(\mathcal{X}^{(1:T)}|\mathcal{X}^{(0)})} \frac{p_{{\boldsymbol{\theta}}}(\mathcal{X}^{(0:T)}|\boldsymbol{z})}{q(\mathcal{X}^{(1:T)}|\mathcal{X}^{(0)})}\right]\\
    &\stackrel{(a)}{\geq} \mathbb{E}_{q(\mathcal{X}^{(1:T)}|\mathcal{X}^{(0)})} \log\left( \frac{p_{{\boldsymbol{\theta}}}(\mathcal{X}^{(0:T)}|\boldsymbol{z})}{q(\mathcal{X}^{(1:T)}|\mathcal{X}^{(0)})} \right) \\
    &\stackrel{(b)}{=}\mathbb{E}_{{q(\mathcal{X}^{(1:T)}|\mathcal{X}^{(0)})}} \left[ \sum_{i=1}^{M}{\log\left( \frac{p_{{\boldsymbol{\theta}}}(\boldsymbol{x}^{(0:T)}_i|\boldsymbol{z})}{q(\boldsymbol{x}^{(1:T)}_i|\boldsymbol{x}^{(0)}_i)} \right)} \right],  \label{eq:diffusion_ELBO} 
\end{align}
where $(a)$ is due to Jensen's inequality, and $(b)$ comes from the independence of the points $\{\boldsymbol{x}^{(t)}_i | i = 1, \cdots, M\}$ that constitute the point cloud $\mathcal{X}^{(t)}$ in each timestep $t$.
This ELBO constitutes the optimization objective for the diffusion model component. 
In the following subsection, we will derive the complete loss function of the proposed Gen-MV sensing model based on this formulation.

\vspace{-0.2cm}
\subsection{Training Objective of the Gen-MV Sensing Model} \label{subsec:training_objective}
The end-to-end training objective of the complete Gen-MV sensing model is to maximize $\mathbb{E}_{q_{\mathrm{data}}(\mathcal{X}^{(0)}, \mathcal{H})}\left[\log{p_{\boldsymbol{\theta}}(\mathcal{X}^{(0)}|\mathcal{H})}\right]$ on paired target point clouds and multi-view channel data $(\mathcal{X}^{(0)}, \mathcal{H})$, where $q_{\mathrm{data}}$ denotes their joint distribution in the dataset.  
Unlike the prior work \cite{xing2025conference} that simply treats the encoder as a trainable module embedded in the diffusion model, 
we derive the ELBO loss $L_{\mathrm{ELBO}}$ of the complete model based on the Gen-MV sensing framework proposed in Section \ref{subsec:gen_framework}, formulated as
\vspace{-0.05cm}
{
\setlength{\belowdisplayskip}{-0.02cm} 
\setlength{\belowdisplayshortskip}{-0.02cm} 
\begin{align}
    &\mathbb{E}_{q_{\mathrm{data}}(\mathcal{X}^{(0)}, \mathcal{H})}\left[-\log{p_{\boldsymbol{\theta}}(\mathcal{X}^{(0)}|\mathcal{H})}\right] \\
    &\stackrel{(a)}{\lessapprox} \mathbb{E}_{q_{\mathrm{data}}(\mathcal{X}^{(0)}, \mathcal{H})} \bigg\{\mathbb{E}_{q_{\boldsymbol{\varphi}}(\boldsymbol{z}|\mathcal{H})} \left[-\log p_{\boldsymbol{\theta}}(\mathcal{X}^{(0)}|\boldsymbol{z})\right] \nonumber \\
    &\quad\quad\quad\quad\quad\quad\quad\quad\quad\quad + D_{\mathrm{KL}}\left(q_{\boldsymbol{\varphi}}(\boldsymbol{z}|\mathcal{H}) \parallel p_{\boldsymbol{\phi}}(\boldsymbol{z})\right) \bigg\} \\
    &\stackrel{(b)}{\leq} \mathbb{E}_{(\mathcal{X}^{(0)}, \mathcal{H})} \bigg\{ \mathbb{E}_{\boldsymbol{z}, \mathcal{X}^{(1:T)}} \sum_{i=1}^{M}{\log\left( \frac{q(\boldsymbol{x}^{(1:T)}_i|\boldsymbol{x}^{(0)}_i)}{p_{{\boldsymbol{\theta}}}(\boldsymbol{x}^{(0:T)}_i|\boldsymbol{z})} \right)} \nonumber \\
    &\quad\quad\quad\quad\quad\quad\quad\quad\quad\quad + D_{\mathrm{KL}}\left(q_{\boldsymbol{\varphi}}(\boldsymbol{z}|\mathcal{H}) \parallel p_{\boldsymbol{\phi}}(\boldsymbol{z})\right) \bigg\} \\
    &\triangleq L_{\mathrm{ELBO}}, 
\end{align}
}where $(a)$ comes from the simplified ELBO loss $\tilde{{L}}_{\mathrm{ELBO}}(\boldsymbol{x}, \mathcal{H})$ of the conditional generative model proposed in (\ref{eq:L_ELBO_relaxed}) with $\boldsymbol{x}$ replaced by $\mathcal{X}^{(0)}$, and $(b)$ is obtained by substituting (\ref{eq:diffusion_ELBO}).
By applying the Markov chain property of the diffusion process specified in (\ref{eq:forward_Markov}) and (\ref{eq:reverse_Markov}) and utilizing Bayes' theorem, 
we can further simplify $L_{\mathrm{ELBO}}$ as 
\vspace{-0.1cm}
{
\setlength{\belowdisplayskip}{-0.02cm} 
\setlength{\belowdisplayshortskip}{-0.02cm} 
\begin{align}
    &L_\text{ELBO} = \mathbb{E}_q \Big[ \sum_{i=1}^{M}
        \underbrace{D_\text{KL}(q(\boldsymbol{x}^{(T)}_i | \boldsymbol{x}^{(0)}_i) \parallel p(\boldsymbol{x}^{(T)}_i))}_{L^{(T)}_i} \nonumber \\[-1.5ex]
        & + \sum_{t=2}^T \underbrace{D_\text{KL}(q(\boldsymbol{x}^{(t-1)}_i | \boldsymbol{x}^{(t)}_i, \boldsymbol{x}^{(0)}_i) \parallel p_{\boldsymbol{\theta}}(\boldsymbol{x}^{(t-1)}_i |\boldsymbol{x}^{(t)}_i, \boldsymbol{z}))}_{L^{(t-1)}_i} \nonumber \\[-1ex]
        & \underbrace{- \log p_{\boldsymbol{\theta}}(\boldsymbol{x}^{(0)}_i | \boldsymbol{x}^{(1)}_i, \boldsymbol{z})}_{L^{(0)}_i}
        + \underbrace{D_{\mathrm{KL}}\left(q_{\boldsymbol{\varphi}}(\boldsymbol{z}|\mathcal{H}) \parallel p_{\boldsymbol{\phi}}(\boldsymbol{z})\right)}_{L_{\mathrm{z}}} \label{eq:L_ELBO_complete}
    \Big],
\end{align}
}where $L^{(T)}_i$ is the KL divergence between two deterministic Gaussian distributions and is independent of learnable parameters. 
$L^{(t-1)}_i$ can be derived from (\ref{eq:reverse_one_step_x0}) and (\ref{eq:reverse_one_step}), which is
\vspace{-0.05cm}
{
\setlength{\belowdisplayskip}{-0.01cm} 
\setlength{\belowdisplayshortskip}{-0.01cm} 
\begin{align} \label{eq: L_t-1}
    \addtolength{\belowdisplayskip}{-0.5ex}
    L^{(t-1)}_i &= \frac{1}{2\tilde{\beta}_t}\left\|\tilde{\boldsymbol{\mu}}_t(\boldsymbol{x}^{(t)}_i,\boldsymbol{x}^{(0)}_i)-\boldsymbol{\mu}_{\boldsymbol{\theta}}(\boldsymbol{x}^{(t)}_i,t,\boldsymbol{z})\right\|^2 \\[-0.5ex]
    &= \frac{\beta_t^2}{2\tilde{\beta}_t \cdot \alpha_t (1-\bar{\alpha}_t)} 
    \left\|\boldsymbol{\epsilon}^{(t)}_i-\boldsymbol{\epsilon}_{\boldsymbol{\theta}}(\boldsymbol{x}_{i}^{(t)},t,\boldsymbol{z})\right\|^2.  \label{eq:L_t-1_i}  
\end{align}
}$L^{(0)}_i$ can be transformed into the same form as $L^{(t-1)}_i$ and merged with it. 
$L_{\mathrm{z}}$ can be regarded as a regularization term that aligns all posterior $q_{\boldsymbol{\varphi}}(\boldsymbol{z}|\mathcal{H})$ from the multi-view encoder to a unified prior $p_{\boldsymbol{\phi}}(\boldsymbol{z})$, 
thereby improving the structural coherence and manifold smoothness of the latent space. 
Here, we employ a normalizing flow \cite{rezende2015NormalizingFlows} parameterized by $\boldsymbol{\phi}$ to model the prior distribution $p_{\boldsymbol{\phi}}(\boldsymbol{z})$,  
which maps a standard Gaussian distribution $p_{\boldsymbol{w}}(\boldsymbol{w}) = \mathcal{N}(\boldsymbol{w}; \boldsymbol{0}, \boldsymbol{I})$ to a complex distribution $p_{\boldsymbol{\phi}}(\boldsymbol{z})$ through a learnable bijective function $F_{\boldsymbol{\phi}}$. 
According to the properties of the flow model, $L_{\mathrm{z}}$ can be formulated as
\vspace{-0.05cm}
\begin{equation} \label{eq:L_z_flow}
\vspace{-0.05cm}
    L_{\mathrm{z}} = D_{\mathrm{KL}}\bigg(q_{\boldsymbol{\varphi}}(\boldsymbol{z}|\mathcal{H})\parallel p_{\boldsymbol{w}}(\boldsymbol{w})\cdot\left|\det{\frac{\partial F_{\boldsymbol{\phi}}}{\partial \boldsymbol{w}}}\right|^{-1}\bigg),   
\end{equation}
where $\boldsymbol{w} = F_{\boldsymbol{\phi}}^{-1}(\boldsymbol{z})$ and $F_{\boldsymbol{\phi}}$ is implemented by the affine coupling layers \cite{dinh2017RealNVP}. 
By randomly sampling timestep $t$ and discarding the timestep-specific coefficients in (\ref{eq:L_t-1_i}), we can reformulate (\ref{eq:L_ELBO_complete}) as the following standard diffusion loss: 
\vspace{-0.1cm}
\begin{equation} \label{eq:loss_standard}
\vspace{-0.1cm}
\begin{split}
    L_{\mathrm{standard}} &= \mathbb{E}_{(\mathcal{H},\mathcal{X}^{(0)}),t,\boldsymbol{z},\boldsymbol{\epsilon}_i} \bigg\{ \frac{1}{4M} \times \\[-1ex]
    &\quad \sum_{i=1}^{M} \left\|\boldsymbol{\epsilon}_{i} - \boldsymbol{\epsilon}_{{\boldsymbol{\theta}}}(\boldsymbol{x}_{i}^{(t)}, t, \boldsymbol{z})\right\|^2 + \gamma_{\mathrm{z}} \cdot L_{\mathrm{z}} \bigg\},
\end{split}
\end{equation}
where averaging is performed over the quantity and dimensionality of points, and a coefficient $\gamma_{\mathrm{z}}$ is introduced to adjust the weight of $L_{\mathrm{z}}$ term. 

In most scenarios, the spatial distribution complexity of targets typically differs between their geometric shape and EM material properties, leading to imbalanced reconstruction challenges associated with these two physical attributes. 
To further enhance reconstruction quality, we introduce the following shape-EM weighted diffusion loss:
\vspace{-0.1cm}
\begin{equation} \label{eq:loss_shape-EM}
\vspace{-0.1cm}
\begin{split}
    L_{\mathrm{shape-EM}} &= \mathbb{E}_{(\mathcal{H},\mathcal{X}^{(0)}),t,\boldsymbol{z},\boldsymbol{\epsilon}_i} \bigg\{ \frac{1}{M} \times \\[-1ex] 
    &\sum_{i=1}^{M} \left(\gamma_{\mathrm{s}} \cdot L_{\mathrm{s}, i} + \gamma_{\mathrm{EM}} \cdot L_{\mathrm{EM}, i} \right) + \gamma_{\mathrm{z}} \cdot L_{\mathrm{z}} \bigg\}, 
\end{split}
\end{equation}
where $L_{\mathrm{s}, i} = \|\boldsymbol{\epsilon}_{i, \mathrm{s}} - \boldsymbol{\epsilon}_{{\boldsymbol{\theta}}, \mathrm{s}}(\boldsymbol{x}_{i}^{(t)}, t, \boldsymbol{z})\|^2$ and  
$L_{\mathrm{EM}, i} = \|\boldsymbol{\epsilon}_{i, \mathrm{EM}} - \boldsymbol{\epsilon}_{{\boldsymbol{\theta}}, \mathrm{EM}}(\boldsymbol{x}_{i}^{(t)}, t, \boldsymbol{z})\|^2$ 
respectively denote the diffusion loss for the $i$-th point in shape and EM dimensions, with subscripts $\mathrm{s}$ and $\mathrm{EM}$ representing the corresponding vector dimensions (i.e. dimensions 1-2 for shape and 3-4 for EM properties).

In the inference stage, we can perform iterative sampling based on (\ref{eq:reverse_Markov})-(\ref{eq:mu_theta}) to generate target point clouds from noise samples. 
The training and inference procedures for the proposed Gen-MV sensing scheme are summarized in Algorithms \ref{algo:train} and \ref{algo:infer}. 

{\color{black}
\subsection{Computational Complexity}
The computation process of the Gen-MV sensing model mainly consists of two components: 
encoding the multi-view channel $\mathcal{H}$ into the target latent code $\boldsymbol{z}$, 
and decoding the reconstructed point cloud $\mathcal{X}^{(0)}$ through the diffusion model.

The computational complexity of multi-view channel encoding mainly comes from the encoder $E$. 
We assume that the hidden layer dimensions of all models are equal to the channel vector dimension $d_{\mathrm{v}}$, and use $L$ to denote the number of model layers. 
Both VS-MLP and MV-BiLSTM have a computational complexity of $O(L BU d_{\mathrm{v}}^2)$. 
The difference lies in that VS-MLP allows parallel processing across views, while MV-BiLSTM performs bidirectional sequential operations and is not parallelizable.  
The computational complexity of MVT is $O(L(B^2 U^2 d_{\mathrm{v}} + BU d_{\mathrm{v}}^2))$, and that of IVT is $O(L(BU^2 d_{\mathrm{v}}^2 + B^2U d_{\mathrm{v}}^2 + BU d_{\mathrm{v}}^2))$. 
IVT exhibits certain advantages over MVT when dealing with a large number of views. Due to the attention mechanism, both MVT and IVT support parallel computation. 

Regarding the diffusion model, its primary computational complexity stems from the noise prediction network $\boldsymbol{\epsilon}_{\boldsymbol{\theta}}$. 
Let $C_{\boldsymbol{\epsilon}}$ denote the computational cost for each point in the point cloud, then according to Algorithm \ref{algo:infer}, 
the computational complexity of the diffusion model in the inference stage is $O(T M C_{\boldsymbol{\epsilon}})$, where $T$ is the number of diffusion steps and $M$ is the number of points. 
According to the network architecture of $\boldsymbol{\epsilon}_{\boldsymbol{\theta}}$ described in Section \ref{subsubsec:reverse_diffusion}, its computational complexity can be denoted as 
\begin{equation}
    C_{\boldsymbol{\epsilon}} = O\left(\sum_{\ell=1}^{L_{\mathrm{d}}} h_{\mathrm{d}}^{(\ell-1)} h_{\mathrm{d}}^{(\ell)} + (d_{\mathrm{t}} + d_{\mathrm{z}}) h_{\mathrm{d}}^{(\ell)} \right), 
\end{equation}
where $h_{\mathrm{d}}^{(\ell)}$ is the dimension of the $\ell$-th ConcatSquash layer's output, and $d_{\mathrm{z}}$ is the dimension of the latent code $\boldsymbol{z}$. 

\begin{algorithm}[!t]
\caption{Training of the Proposed Gen-MV Sensing}
\label{algo:train}
\begin{algorithmic}[1]
\Repeat
    \ParState{Sample multi-view channel data and the corresponding target point cloud $(\mathcal{H},\mathcal{X}^{(0)})$ from the training set.}
    \ParState{Sample target latent code $\boldsymbol{z} \sim q_{\boldsymbol{\varphi}}(\boldsymbol{z}|\mathcal{H})$ through the multi-view channel encoder $\boldsymbol{\varphi}$.}
    
    \State Sample $t \sim \text{Uniform}(\{1, \dots, T\})$.
    \For{$i=1$ to $M$ in parallel}
        \State Sample $\boldsymbol{\epsilon}_i \sim \mathcal{N}(\boldsymbol{0}, \mathbf{I})$.
        \State $\boldsymbol{x}_i^{(t)}=\sqrt{\bar{\alpha}_t}\boldsymbol{x}_i^{(0)} + \sqrt{1-\bar{\alpha}_t} \boldsymbol{\epsilon}_i$.
        \State $L_{\mathrm{s}, i} = \|\boldsymbol{\epsilon}_{i, \mathrm{s}} - \boldsymbol{\epsilon}_{{\boldsymbol{\theta}}, \mathrm{s}}(\boldsymbol{x}_{i}^{(t)}, t, \boldsymbol{z})\|^2$.
        \State $L_{\mathrm{EM}, i} = \|\boldsymbol{\epsilon}_{i, \mathrm{EM}} - \boldsymbol{\epsilon}_{{\boldsymbol{\theta}}, \mathrm{EM}}(\boldsymbol{x}_{i}^{(t)}, t, \boldsymbol{z})\|^2$.
    \EndFor
    \State $L_{\mathrm{z}} = D_{\mathrm{KL}}\big(q_{\boldsymbol{\varphi}}(\boldsymbol{z}|\mathcal{H})\parallel p_{\boldsymbol{w}}(F_{\boldsymbol{\phi}}^{-1}(\boldsymbol{z}))\cdot\left|\det{\frac{\partial F_{\boldsymbol{\phi}}}{\partial \boldsymbol{w}}}\right|^{-1}\big)$.
    \State Take gradient descent step on:
    \Statex \quad \hfill $\nabla_{\boldsymbol{\varphi},\boldsymbol{\phi},\boldsymbol{\theta}} \frac{1}{M} \sum_{i=1}^{M} \left(\gamma_{\mathrm{s}} \cdot L_{\mathrm{s}, i} + \gamma_{\mathrm{EM}} \cdot L_{\mathrm{EM}, i} \right) + \gamma_{\mathrm{z}} \cdot L_{\mathrm{z}}.$ \hfill \null
\Until{converged}
\end{algorithmic}
\end{algorithm}

\section{Numerical Results} \label{sec:numerical_results}
In this section, we evaluate the performance of the proposed Gen-MV sensing scheme. 
First, we introduce the experimental parameter settings and define the performance evaluation metric. 
Then, we present a variety of experimental results to comprehensively evaluate the models' sensing performance. 
Finally, we conduct ablation studies to validate the effectiveness of the positional embedding and the shape-EM weighted diffusion loss in our proposed scheme. 

\subsection{Experiment Settings} \label{subsec:exp_settings}
In this work, we set the RoI as a 2-D square region centered at $(0,0)$ with dimensions $0.5\mathrm{m}\times 0.5\mathrm{m}$. 
The system center frequency $f_c = 3 \mathrm{GHz}$, and we use $N_c=8$ subcarriers to transmit pilots with a subcarrier spacing $\Delta f=100 \mathrm{kHz}$. 
The BSs are deployed within a distance $R_r$ of $80$m to $100$m from the origin, each equipped with $N_r=4$ antennas in a ULA configuration, with the array normal pointing toward the origin. The total number of BSs $B\leq 16$. 
The UEs are randomly distributed at a distance $R_t$ of $4$m to $10$m from the origin, with the number of UEs $U\leq 32$. 
In practice, this can refer to snapshots of several UEs moving within this area. 
By default, we synthesize homogeneous targets based on the MNIST dataset \cite{lecun1998MNIST}, assigning each target a relative permittivity $\varepsilon_r\in [1.5,2.5]$ and a conductivity $\sigma\in[0,0.1]$ (S/m). 
Certain experiments use different dataset settings, as detailed in their corresponding descriptions. 
The scattering channel is simulated using the MoM, as described in Section \ref{subsec:channel_model}. 
We randomly generate 50,000 data samples, which are divided into training, validation, and test sets in an 8:1:1 ratio. 

\begin{algorithm}[!t]
\caption{Inference of the Proposed Gen-MV Sensing} \label{algo:infer}
\begin{algorithmic}[1]
\Require Multi-view channel data $\mathcal{H}$, including each view's CSI and BS/UE positions.
\Ensure Reconstructed target point cloud $\hat{\mathcal{X}}^{(0)}$.
\State Sample target latent code $\boldsymbol{z} \sim q_{\boldsymbol{\varphi}}(\boldsymbol{z}|\mathcal{H})$ through the multi-view channel encoder $\boldsymbol{\varphi}$.
\For{$i=1$ to $M$ in parallel}
    \State Sample $\hat{\boldsymbol{x}}^{(T)}_i \sim \mathcal{N}(\boldsymbol{0},\mathbf{I})$.
    \For{$t=T$ to $1$}
        \State $\hat{\boldsymbol{\epsilon}}^{(t)}_{i} = \boldsymbol{\epsilon}_{{\boldsymbol{\theta}}}(\hat{\boldsymbol{x}}^{(t)}_i, t, \boldsymbol{z})$.
        \State Sample $\boldsymbol{\epsilon}\sim\mathcal{N}(\boldsymbol{0}, \mathbf{I})$.
        \State $\hat{\boldsymbol{x}}^{(t-1)}_{i} = \frac{1}{\sqrt{\alpha_t}} \left( \hat{\boldsymbol{x}}^{(t)}_{i} - \frac{1 - \alpha_t}{\sqrt{1 - \bar{\alpha}_t}} \hat{\boldsymbol{\epsilon}}^{(t)}_{i} \right) + \tilde{\beta}_t\cdot\boldsymbol{\epsilon}$.
    \EndFor
\EndFor
\State $\hat{\mathcal{X}}^{(0)} = \{\hat{\boldsymbol{x}}^{(0)}_{i}|i=1,\cdots,M\}$.
\end{algorithmic}
\end{algorithm}
}

To evaluate the impact of communication link quality and pilot resources on sensing performance, 
we adopt a basic least squares (LS) channel estimation scheme as a case study. 
In the uplink scenario considered in this paper, the pilot signal transmitted by the $u$-th UE on the $n$-th subcarrier and received by the $b$-th BS is denoted by 
$\mathbf{Y}_{b,u,n} = \mathbf{h}_{b,u,n} \mathbf{s}_{b,u,n}^{\mathsf{T}} + \mathbf{Z}_{b,u,n}$, 
where $\mathbf{s}_{b,u,n}\in\mathbb{C}^{L\times 1}$ represents the transmitted pilot signal and $L$ is the number of pilot symbols. 
The channel $\mathbf{h}_{b,u,n}\in\mathbb{C}^{N_r\times 1}$ is defined in (\ref{eq:SV_spatial_channel}) and $\mathbf{Z}_{b,u,n}$ is additive white complex Gaussian noise. 
The corresponding LS channel estimation result is 
$\hat{\mathbf{h}}_{\mathrm{LS},b,u,n} = \mathbf{Y}_{b,u,n} \mathbf{s}_{b,u,n}^{*} (\mathbf{s}_{b,u,n}^{\mathsf{T}} \mathbf{s}_{b,u,n}^{*})^{-1}$. 
Based on (\ref{eq:SV_spatial_channel})-(\ref{eq:H_set}), we compile the estimated $\hat{\mathbf{h}}_{\mathrm{LS},b,u,n}$ across all transceiver pairs and subcarrier frequencies into multi-view channel data $\hat{\mathcal{H}}_{\mathrm{LS}}$, 
which serves as the input to the Gen-MV sensing model in the relevant evaluation experiments. 

To demonstrate the advantages of the proposed Gen-MV sensing schemes over existing solutions, 
we introduce two iterative EM imaging algorithms, Born iterative method (BIM) and BIM combined with compressed sensing (BIM-CS), as external baselines, 
with reference to \cite{jiang2024EMpropertySensing} and \cite{jiang2025multi-BS}. 
The detailed descriptions of BIM and BIM-CS can be found in Appendix \ref{app:BIM_BIM-CS}.
To analyze the capability of different network architectures in extracting multi-view features for wireless sensing, 
we construct the multi-view channel encoders using four different structures proposed in Section \ref{subsec:MV_encoder} for experimental comparison. 
Among them, VS-MLP, MVT, and IVT consist of 6 network layers, while MV-BiLSTM employs 3 layers of LSTM in each direction. 
Other model parameters remain consistent, with configurations detailed in Table \ref{tab:model_settings}.
The noise prediction network $\boldsymbol{\epsilon}_{\boldsymbol{\theta}}$ uses ConcatSquash layers with dimensions 4-128-256-512-256-128-4, 
and the timestep encoding dimension $d_{\mathrm{t}}=32$. 
The weighting coefficients in the loss function (\ref{eq:loss_shape-EM}) are set as $\gamma_{\mathrm{s}}=0.45$, $\gamma_{\mathrm{EM}}=0.05$, and $\gamma_{\mathrm{z}}=1\times10^{-4}$. 
Under the settings of $B=16$ and $U=32$, the number of parameters and floating-point operations (FLOPs) of the four multi-view channel encoders are summarized in Table \ref{tab:Params_and_FLOPs}.
During training, we set the batch size to 128, the initial learning rate to $10^{-4}$, and decay the learning rate by a factor of 0.8 every $10^5$ steps. 
The total training process takes $10^6$ steps, and the Adam optimizer is used for optimization. 

\begin{table}[!t]
  \centering
  \caption{Model parameter settings} \label{tab:model_settings}
  \begin{tabular}{lp{3cm}} 
    \toprule
    \textbf{Parameters} & \textbf{Value} \\ 
    \midrule 
    Number of encoding frequencies $d_{\mathrm{p}}$   &  10 \\
    Channel feature dimension $d_{\mathrm{v}}$   &  256 \\
    Target latent code dimension $d_{\mathrm{z}}$   &  128 \\
    Number of points $M$ in the point cloud  &  1000 \\
    Maximum diffusion timestep $T$   &  100 \\
    Variance schedule $\{\beta_t\}$  &  $\beta_1=1\times10^{-4}$, linearly increasing to $\beta_T=0.02$. \\
    \bottomrule
  \end{tabular}
\end{table}

\begin{table}[!t]
  \centering
  \caption{Parameter counts and FLOPs of four multi-view channel encoder architectures} \label{tab:Params_and_FLOPs}
  \begin{tabular}{p{2cm}p{2.5cm}p{1.5cm}} 
    \toprule
    \textbf{Model} & \textbf{Parameter Counts} & \textbf{FLOPs} \\ 
    \midrule 
    VS-MLP  &  1.59 Million  &  1.52 Giga \\
    MV-BiLSTM  &  4.87 Million  &  4.88 Giga \\
    MVT   &  3.30 Million  &  4.91 Giga \\
    IVT   &  4.88 Million  &  5.06 Giga \\
    \bottomrule
  \end{tabular}
\vspace{-0.1cm}
\end{table}

To comprehensively evaluate the models' reconstruction performance in both shape and EM properties, 
we adopt the log-scale Chamfer Distance (log-CD) to evaluate target point cloud reconstruction quality, defined as
\begin{equation} \label{eq:log-CD}
\begin{split}
    \text{log-CD}\mathrm{~(dB)} = 10 \log_{10} \bigg( 
        & \frac{1}{M}\sum_{\boldsymbol{x}\in\mathcal{X}}\min_{\hat{\boldsymbol{x}}\in\hat{\mathcal{X}}}\|\boldsymbol{x}-\hat{\boldsymbol{x}}\|_2^2 \\
        & + \frac{1}{M}\sum_{\hat{\boldsymbol{x}}\in\hat{\mathcal{X}}}\min_{\boldsymbol{x}\in\mathcal{X}}\|\hat{\boldsymbol{x}}-\boldsymbol{x}\|_2^2
    \bigg), 
\end{split}
\end{equation}
where $\mathcal{X}$ is the point cloud sampled from the ground truth target and $\hat{\mathcal{X}}$ is the corresponding reconstruction result, 
and the logarithmic operation is applied to enhance the metric's sensitivity to small numerical differences. 
All evaluations are performed on dimensionless normalized shape-EM point clouds as defined in (\ref{eq:normalized_point}). 
Unless otherwise specified, the default number of views is set to $B=16$ and $U=32$ in performance evaluations to assess the performance limits of the models.

\vspace{-0.2cm}

\subsection{Performance Evaluation}  \label{subsec:performance_eval}
\subsubsection{Generative Reconstruction Process}
\begin{figure}[!htbp] 
\centering
\includegraphics[width=0.95\linewidth]{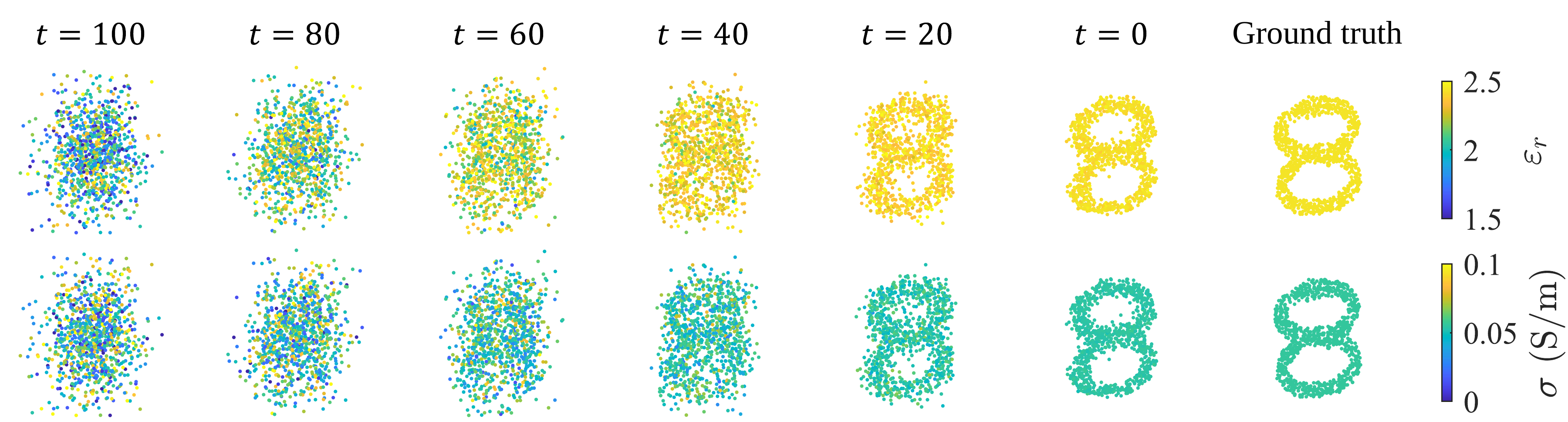}
\caption{Generative reconstruction process of a target based on the conditional point cloud diffusion model, 
where the multi-view channel encoder is implemented with IVT structure.}
\label{fig:diffusion_process}
\vspace{-0.2 cm}
\end{figure}

Fig. \ref{fig:diffusion_process} visualizes the generative reconstruction process of the conditional point cloud diffusion model for a target from the test set, 
where an IVT-based encoder is implemented to process multi-view channel data $\mathcal{H}$. 
As proposed in our scheme, the target latent code $\boldsymbol{z}$, which is extracted from $\mathcal{H}$ by the multi-view channel encoder, 
guides the diffusion model to progressively generate high-quality target point clouds from random noise distributions, 
including clear geometric shapes and accurate EM property estimations. 

\begin{figure}[!htbp] 
\vspace{-0.2cm}
\centering
\includegraphics[width=0.75\linewidth]{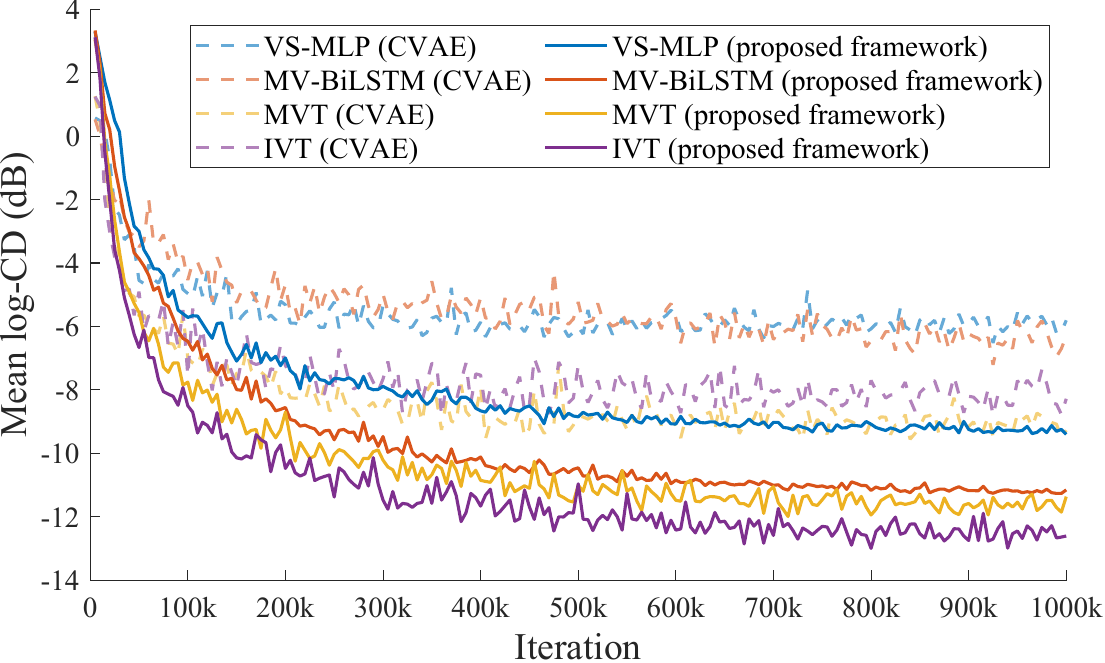}
\caption{\color{black}Validation performance during training for models using four multi-view channel encoders 
under both the standard CVAE and the proposed simplified framework.}
\label{fig:converge_4algo_and_CVAE}
\vspace{-0.2 cm}
\end{figure}

{\color{black}
\subsubsection{Efficiency of the Gen-MV Sensing Framework} 
Fig. \ref{fig:converge_4algo_and_CVAE} shows the training-time validation performance of the proposed Gen-MV sensing models, 
with four multi-view channel encoders implemented under both the standard CVAE and the simplified framework described in Section \ref{subsec:gen_framework}. 
The results show that all models converge to an acceptable level of performance, confirming the feasibility of both approaches. 
However, compared to the proposed framework, the standard CVAE leads to significant performance degradation and more unstable convergence behavior across all cases. 
These experimental results verify that the simplified conditional generation framework achieves better performance on Gen-MV sensing with lower model complexity and computational cost. 
}

\begin{figure*}[!t]
    \centering
    \begin{subfigure}[t]{0.68\linewidth}
        \centering
        \includegraphics[width=\linewidth]{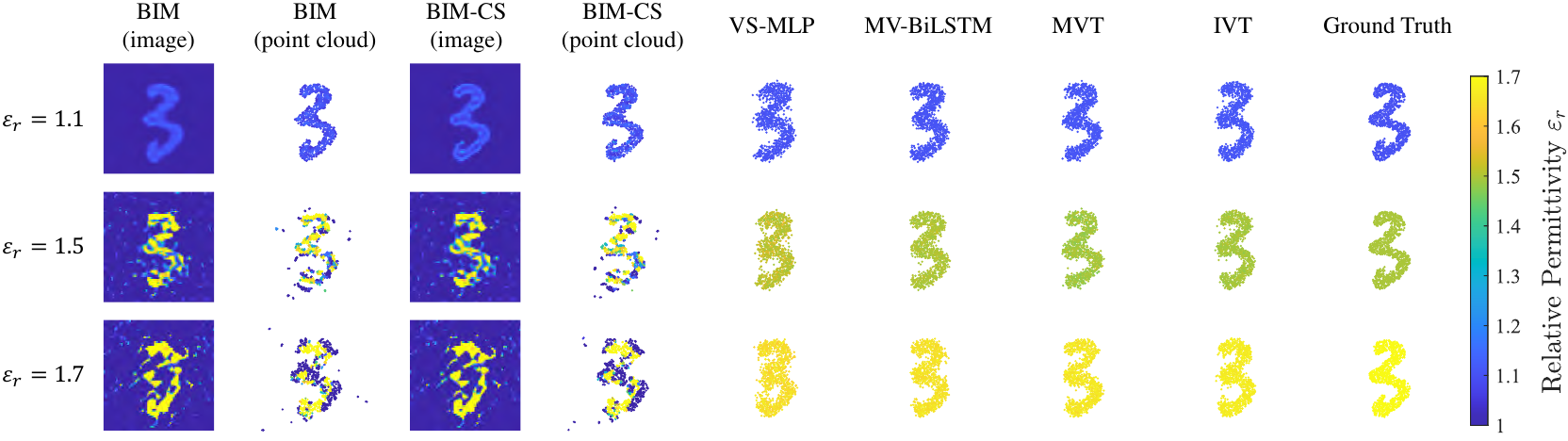}
        \caption{Relative permittivity.}
        \label{fig:show_example_epsr_1.1to1.7}
    \end{subfigure}
    \vfill
    \hspace*{0.75pt}
    \begin{subfigure}[t]{0.68\linewidth}
        \vspace{0.2cm}
        \centering
        \includegraphics[width=\linewidth]{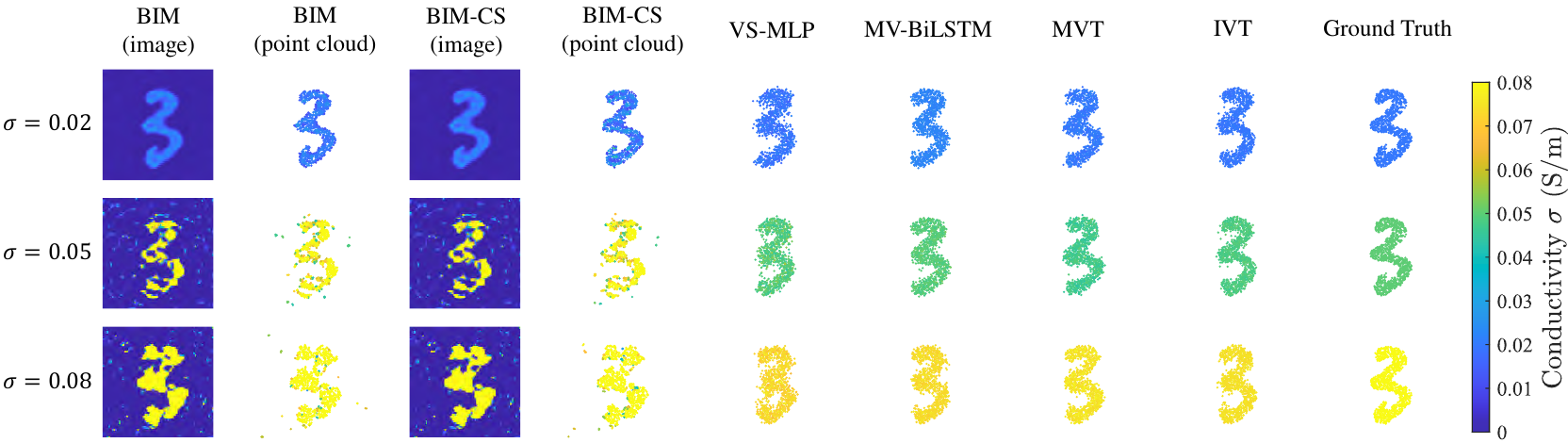}
        \caption{Conductivity}
        \label{fig:show_example_sigma_0to0.08}
    \end{subfigure}
    \vspace{-0.1cm}
    \caption{\color{black}Reconstruction results of BIM, BIM-CS, and the Gen-MV sensing models based on four multi-view channel encoders for targets with different EM properties under the identical system configurations. 
    Among them, BIM and BIM-CS are pixel-based reconstruction algorithms. We use the K-means algorithm to detect the target regions and sample point clouds, 
    with both the original reconstructed images and point clouds displayed in the figure.}
    \label{fig:show_example_low_contrast}
    \vspace{-0.2cm}
\end{figure*}

\begin{figure}[!t] 
\centering
\includegraphics[width=0.85\linewidth]{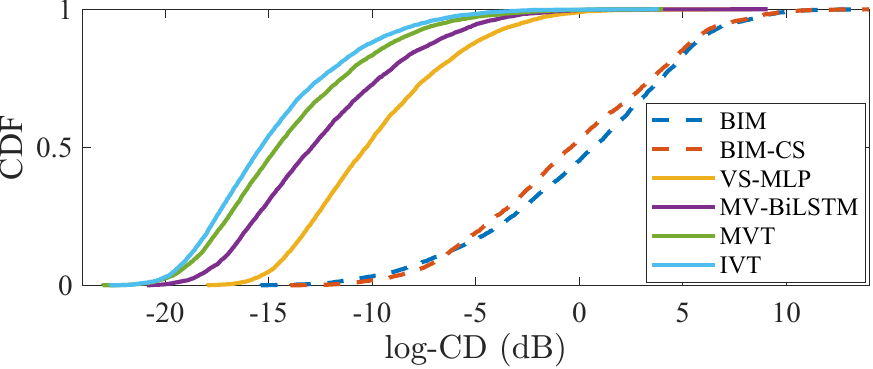}
\hspace{0.1 cm}\null
\caption{\color{black} Performance distribution of BIM, BIM-CS, and the Gen-MV sensing models based on four encoders on the low-contrast target test set.}
\label{fig:chamfer_dist_dB_cdf_6algo}
\vspace{-0.4 cm}
\end{figure}

\begin{figure*}[!t]
\centering
\begin{subfigure}[t]{0.52\linewidth}   
\centering
    \includegraphics[width=\linewidth]{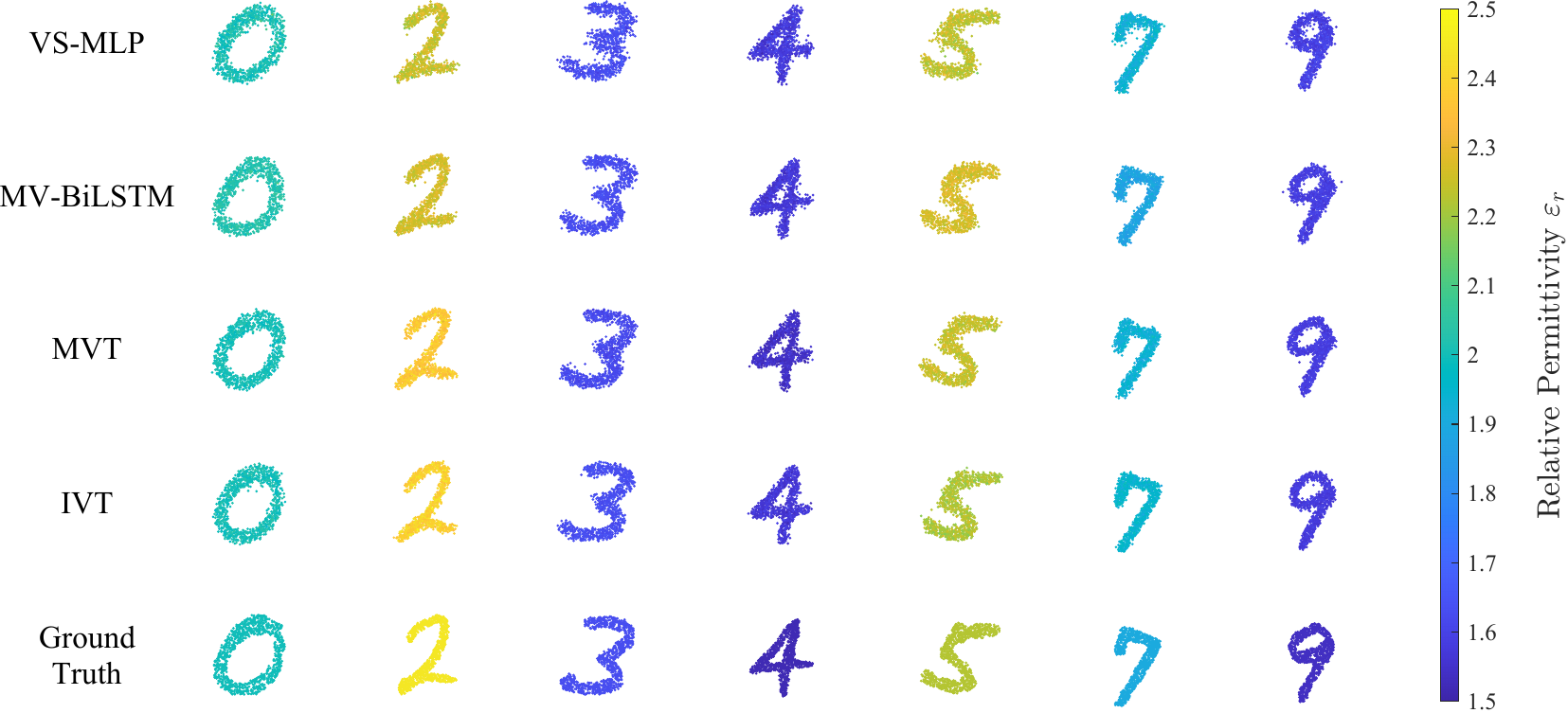}
    \caption{Relative permittivity.}
    \label{fig:show_eample_epsr}
\end{subfigure}
\hfill
\begin{subfigure}[t]{0.45\linewidth}   
\centering
    \includegraphics[width=\linewidth]{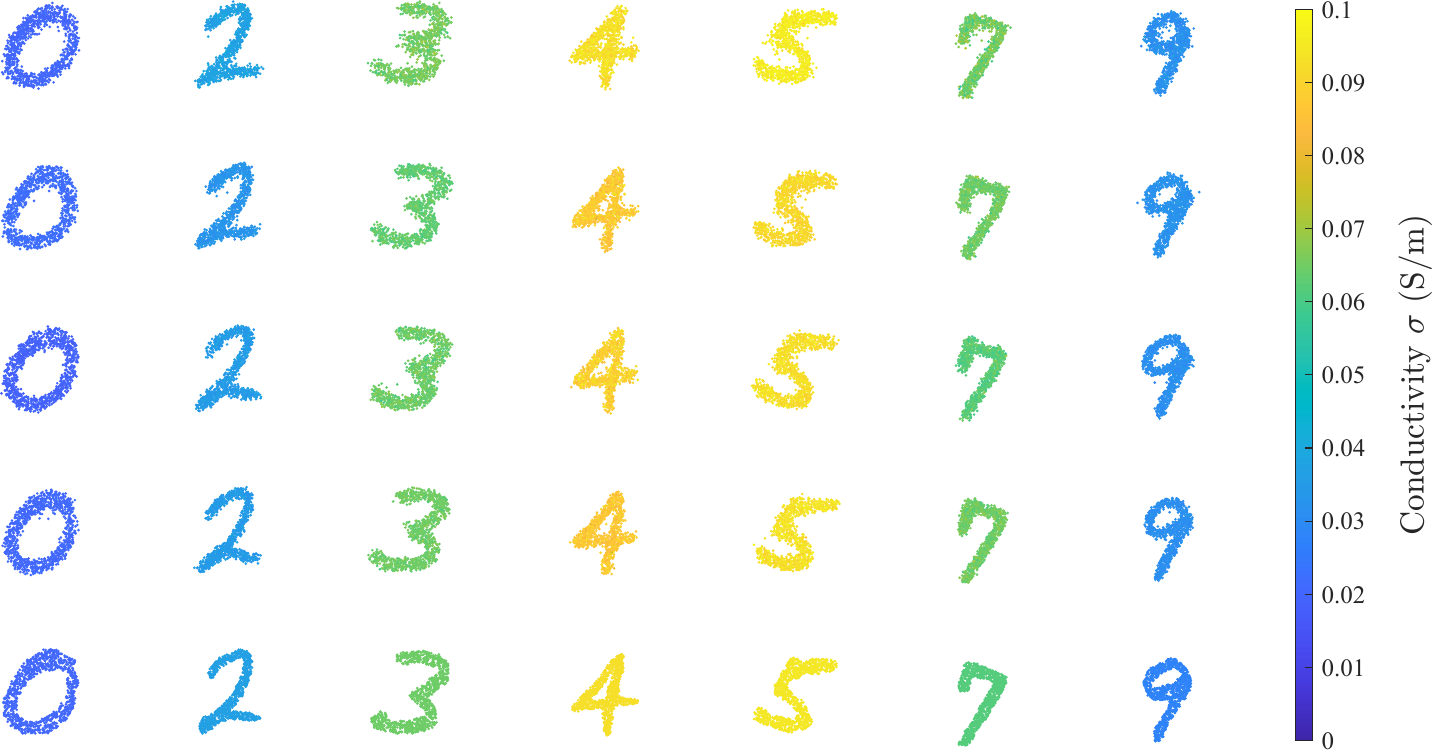}
    \caption{Conductivity.\mbox{~~~~~~~~~~~~~~~}}
    \label{fig:show_eample_sigma}
\end{subfigure}
\hspace{0.1cm}
\caption{Reconstruction results of several targets using four multi-view channel encoder architectures (i.e. VS-MLP, MV-BiLSTM, MVT, and IVT). 
The relative permittivity $\varepsilon_r$ and conductivity $\sigma$ at each point position are visualized by color mapping.}
\label{fig:show_example}
\vspace{-0.2cm}
\end{figure*}

\begin{figure} [!t] 
\centering
\includegraphics[width=0.8\linewidth]{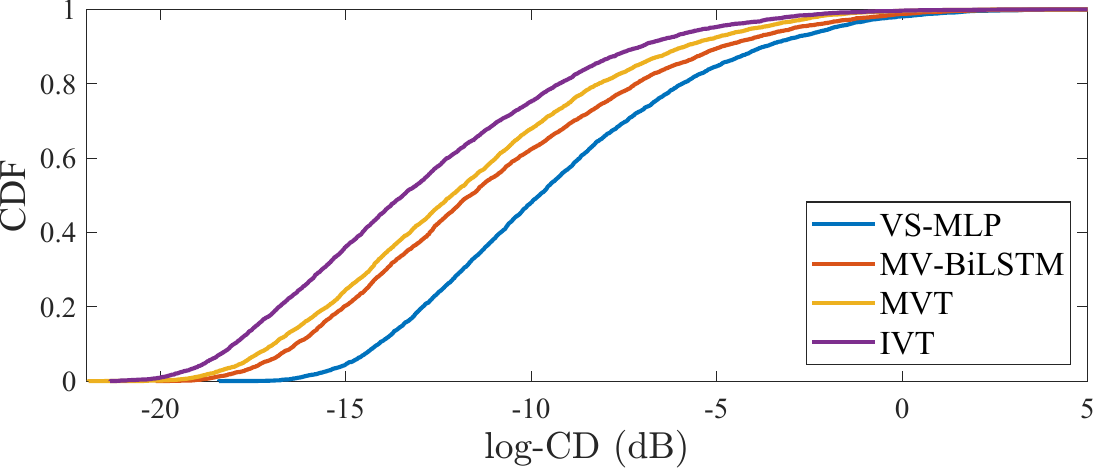}
\hspace{0.2cm}
\caption{CDFs of log-CD on the test set for four multi-view channel encoder architectures.}
\label{fig:cdf}
\vspace{-0.3cm}
\end{figure}

\subsubsection{Performance of the Gen-MV Sensing Schemes and Baseline Methods} 
We first investigate the range of target EM properties applicable to each algorithm. 
Fig. \ref{fig:show_example_low_contrast} visualizes the reconstruction results of all six algorithms for targets with different EM properties $(\varepsilon_r, \sigma)$. 
For the target with $\varepsilon_r=1.1, \sigma=0.02~(\mathrm{S/m})$, both BIM and BIM-CS perform well, and their reconstruction results exhibit higher clarity compared to the Gen-MV sensing methods, 
as BIM and BIM-CS are founded on rigorous physical models. 
However, for the target with $\varepsilon_r=1.5, \sigma=0.05~(\mathrm{S/m})$, the reconstruction quality of BIM and BIM-CS significantly deteriorates, 
exhibiting ripples inside the reconstructed scatterers and artifacts outside, along with noticeable deviation in EM parameter estimation. 
When the target contrast further increases to $\varepsilon_r=1.7, \sigma=0.08~(\mathrm{S/m})$, the reconstruction results of BIM and BIM-CS show more obvious distortion. 
This is because the core of BIM and BIM-CS, the Born iteration, essentially involves repeated applications of the first-order Born approximation \cite{wang1989BIM}. 
It can only perform well under weak scattering conditions, where the target's relative permittivity and conductivity are assumed to be small. 
Due to the sparsity constraint introduced in BIM-CS, it produces fewer artifacts than BIM, but the overall improvement remains limited. 
For targets with even higher contrast, these two baselines tend to diverge. 
In comparison, the four Gen-MV sensing models deliver relatively stable and accurate reconstruction results across various target contrasts. 
This advantage stems from the powerful nonlinear representation capability of generative models, which reduces the models' dependence on weak scattering conditions. 

Next, we conduct performance comparisons for these six algorithms on the test set. 
Given the significant performance degradation of BIM and BIM-CS for high-contrast targets, 
we generate targets within the ranges of $\varepsilon_r\in[1.1, 1.5]$ and $\sigma\in[0, 0.05]$ for training and testing. 
Fig. \ref{fig:chamfer_dist_dB_cdf_6algo} presents the cumulative distribution functions (CDFs) of the log-CD metric on the test set for all algorithms. 
It can be observed that BIM-CS benefits from the sparsity prior and achieves moderate performance gains over BIM on most samples. 
However, even under low-contrast conditions, BIM and BIM-CS still exhibit a significant performance gap compared to the four Gen-MV sensing models proposed in this paper, 
highlighting the substantial advantage of Gen-AI over traditional computational EM imaging methods. 
Among the four encoders, MV-BiLSTM and MVT outperform VS-MLP due to their ability to exchange information across multiple views, 
and IVT achieves the best performance by further utilizing the structural characteristics of multi-view channels. 
Considering the significant performance gap between BIM/BIM-CS and the Gen-MV sensing schemes, 
we exclude them from the remaining experiments in the manuscript and reset the target's EM parameter range to the more challenging region of $\varepsilon_r \in [1.5, 2.5]$ and $\sigma \in [0, 0.1]$.

For the proposed Gen-MV sensing models, the structural design of the multi-view channel encoder determines the effectiveness of extracting target features from multi-view CSI, and thus becomes the performance bottleneck for the subsequent target generation. 
To provide a more intuitive and comprehensive illustration of the encoder's impact on the Gen-MV sensing model, 
Fig. \ref{fig:show_example} visualizes several reconstructed targets of four multi-view encoder architectures (VS-MLP, MV-BiLSTM, MVT, and IVT), 
while Fig. \ref{fig:cdf} shows their performance distributions on the test set. 
It can be observed that the reconstruction results produced by VS-MLP, which processes each view independently, are relatively blurry. 
MV-BiLSTM and MVT significantly improve the model’s sensing performance by introducing multi-view fusion mechanisms. 
IVT further exploits the inherent structure of the multi-view channels through a novel interleaved transmitter-receiver view attention mechanism, 
thereby achieving both a higher sensing performance limit and better results on hard samples. 
The performance differences among the four encoders are also consistent with the results on the low-contrast target dataset shown in Fig. \ref{fig:chamfer_dist_dB_cdf_6algo}. 
These experimental results clearly demonstrate the effectiveness of incorporating multi-view information fusion in the encoder and designing fusion mechanisms that align with the data structure.

\subsubsection{Performance under Different Numbers of Views}
\begin{figure}[!t] 
\vspace{-0.2 cm}
\centering
\includegraphics[width=0.8\linewidth]{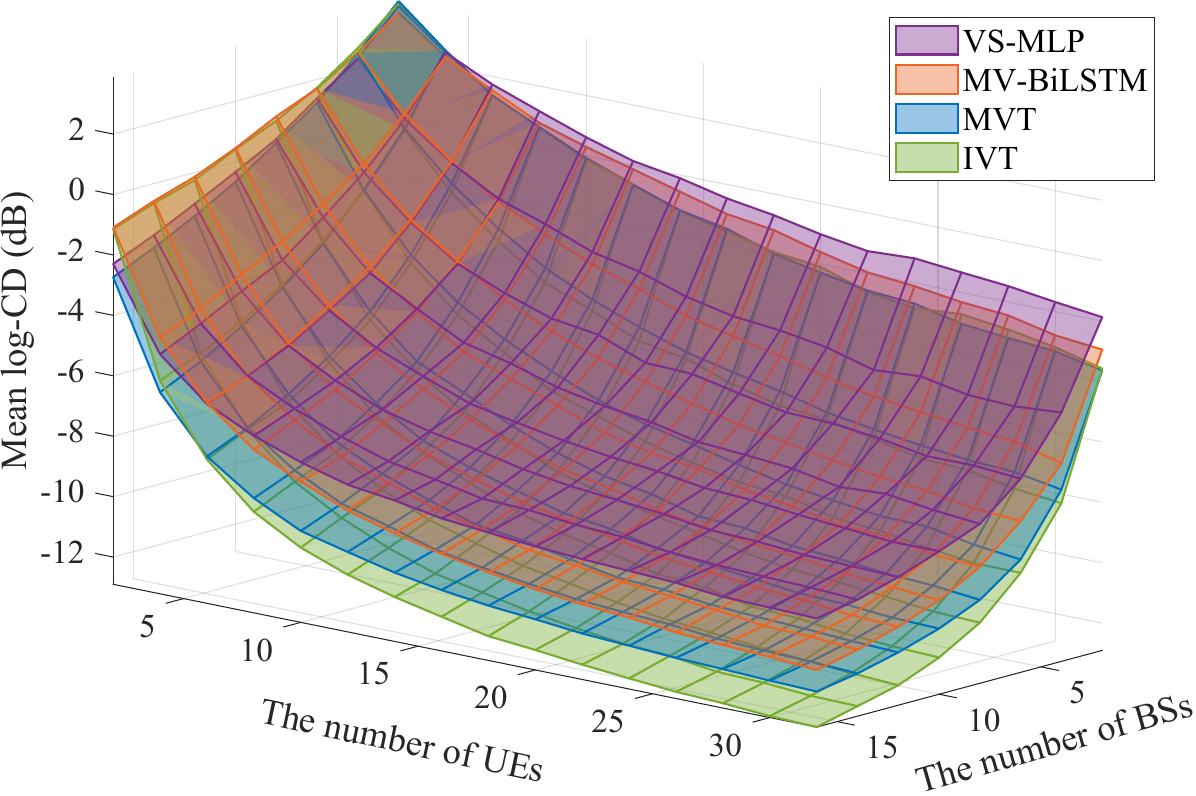}
\hspace{0.3 cm}
\caption{Mean log-CD of reconstruction results for four multi-view channel encoder architectures across different numbers of BSs and UEs.}
\label{fig:surf3D_4algo_view}
\vspace{-0.3cm}
\end{figure}

To analyze the importance of leveraging multi-view channel measurements in wireless sensing and evaluate the capability of different model architectures in processing multi-view data, 
Fig. \ref{fig:surf3D_4algo_view} shows the mean log-CD of reconstruction results for four encoder architectures under variable numbers of BSs and UEs, 
and Table \ref{tab:performance_view} lists the mean log-CD metirc of these four encoders under three different numbers of views. 
It can be observed that all four models can operate effectively with variable BS/UE configurations, and the reconstruction quality is positively correlated with the number of views. 
It demonstrates that the proposed Gen-MV sensing framework can handle flexible BS/UE quantities and positions, 
effectively extracting target features from multi-view CSI to accomplish sensing tasks.
As shown in Table \ref{tab:performance_view}, IVT's performance with 8 BSs and 16 UEs approaches that of VS-MLP with 16 BSs and 32 UEs, 
which highlights the critical role of multi-view feature extraction architectures for the Gen-MV sensing models. 
Although multiple views are beneficial for obtaining comprehensive observations of targets, 
the improvement in reconstruction quality gradually diminishes as the number of views increases. 
This reflects the marginal effect of enhancing sensing accuracy solely by increasing the number of views. 

\begin{table}[!htbp]
  \centering
  \caption{Mean log-CD (dB) of four multi-view channel encoder architectures under different view configurations. The number of views $N_{\mathrm{view}}$ is represented as (number of BSs, number of UEs). } \label{tab:performance_view}
  \begin{tabular}{l|>{\centering}p{1.2cm}>{\centering}p{1.2cm}>{\centering}p{1.2cm}} 
    \toprule
    \diagbox{\makecell{\textbf{model}}}{\makecell{$N_{\mathrm{view}}$}} & \textbf{(4, 8)} & \textbf{(8, 16)} & \textbf{(16, 32)} \\ 
    \midrule 
    VS-MLP     &  -1.37  &  -6.25  &  -9.30 \\
    MV-BiLSTM  &  -1.48  &  -7.14  &  -11.01 \\
    MVT        &  -2.35  &  -8.57  &  -11.72 \\
    IVT        &  \textbf{-2.78}  &  \textbf{-9.16}  &  \textbf{-12.90} \\
    \bottomrule
  \end{tabular}
\end{table}

\subsubsection{Impact of Communication Link Quality, Pilot Resources, and Environmental Clutter on Sensing Performance}
To evaluate the applicability of our models under non-ideal CSI, we analyze how the sensing performance varies with SNR, the number of pilot symbols $L$, and the number of clutter scatterers under the LS channel estimation mentioned in Section \ref{subsec:exp_settings}. 
We randomly generate Quadrature Phase Shift Keying (QPSK)-modulated symbols to simulate the pilot signal. 
The models are trained with $L = 32$ and $\mathrm{SNR}\in[10,20]$ (dB), and tested over a wider range of parameters. 
{If the RCS $\sigma_{\mathrm{rcs}}$ is used to approximate the target scattering intensity, 
then SNR is formulated as 
\begin{equation}
    \text{SNR} = \left(\frac{P_t G_t G_r \lambda^2 \sigma_{\rm{rcs}}}{(4 \pi)^3 R_t^2 R_r^2}\right) / (k T_r B_f), 
\end{equation} 
where the signal bandwidth is $B_f = N_c \Delta f$, the antenna gains are $G_r = G_t = 3$ dBi\cite{jiang2024EMdiff}, the temperature is $T_r = 290\,\text{K}$, and $k = 1.38\times 10^{-23}\,\rm{J/K}$ is the Boltzmann constant. 
When the UE transmit power $P_t$ is limited to 23 dBm, even for a scatterer with an RCS as low as 0.001 m$^2$, the BS can obtain an SNR greater than 0 dB. 
This essentially validates the feasibility of the system settings in this paper.}

\begin{figure*}[!t]
\centering
\hfill
\begin{subfigure}[t]{0.3\linewidth}
    \includegraphics[width=\linewidth]{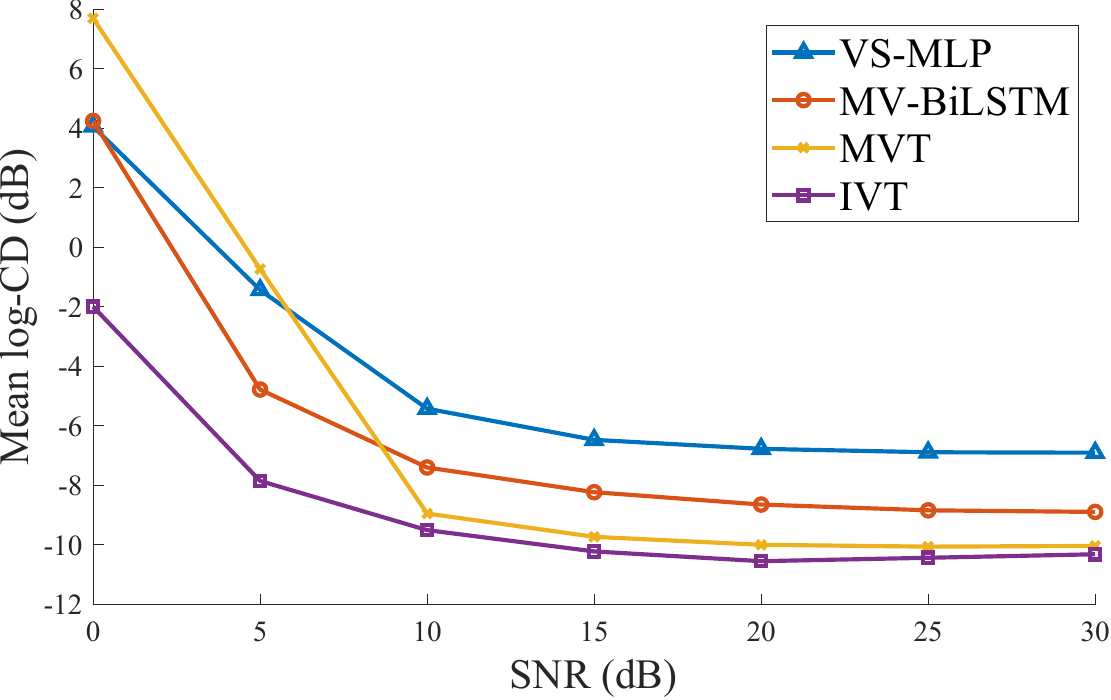}
    \caption{\color{black}Performance of the Gen-MV sensing models with four encoders under different SNR levels.}
    \label{fig:performance_signalSNRdB_Ns32_4algo}
\end{subfigure}
\hfill
\begin{subfigure}[t]{0.3\linewidth}
    \includegraphics[width=\linewidth]{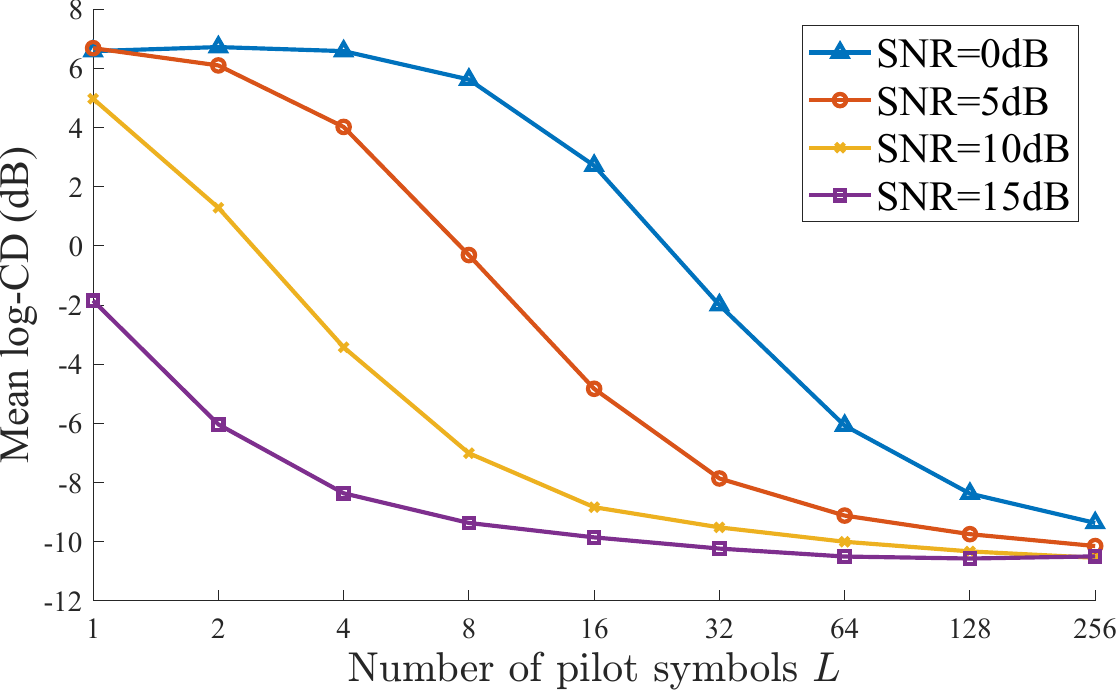}
    \caption{\color{black}Performance of the IVT-based model under different numbers of pilot symbols and SNRs.}
    \label{fig:performance_Nsymbol_differentSNR_IVT}
\end{subfigure}
\hfill
\begin{subfigure}[t]{0.3\linewidth}
    \includegraphics[width=\linewidth]{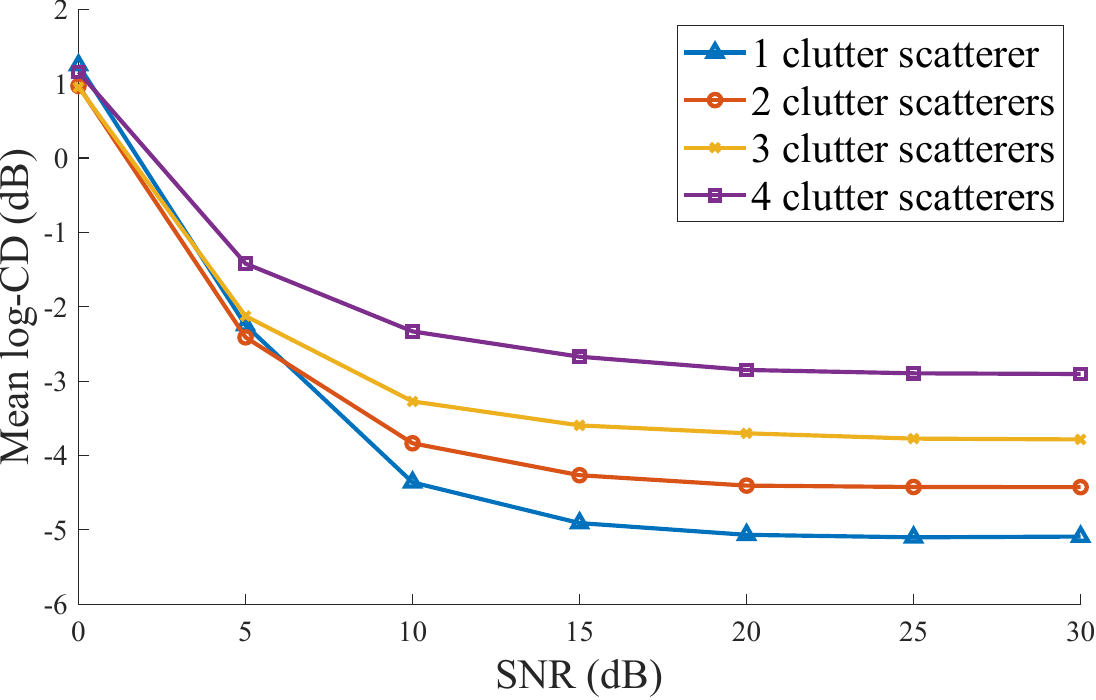}
    \caption{\color{black}Performance of the IVT-based model under different numbers of clutter scatterers and SNRs.}
    \label{fig:performance_clutterN1to4_signalSNRdB_Ns32}
\end{subfigure}
\caption{\color{black} Impact of communication link quality, pilot resources, and environmental clutter on the proposed Gen-MV sensing schemes. }
\label{fig:performance_link}
\vspace{-0.16cm}
\end{figure*}

\begin{figure*}[!t]
\centering
\hfill
\begin{subfigure}[t]{0.25\linewidth}
    \includegraphics[width=\linewidth]{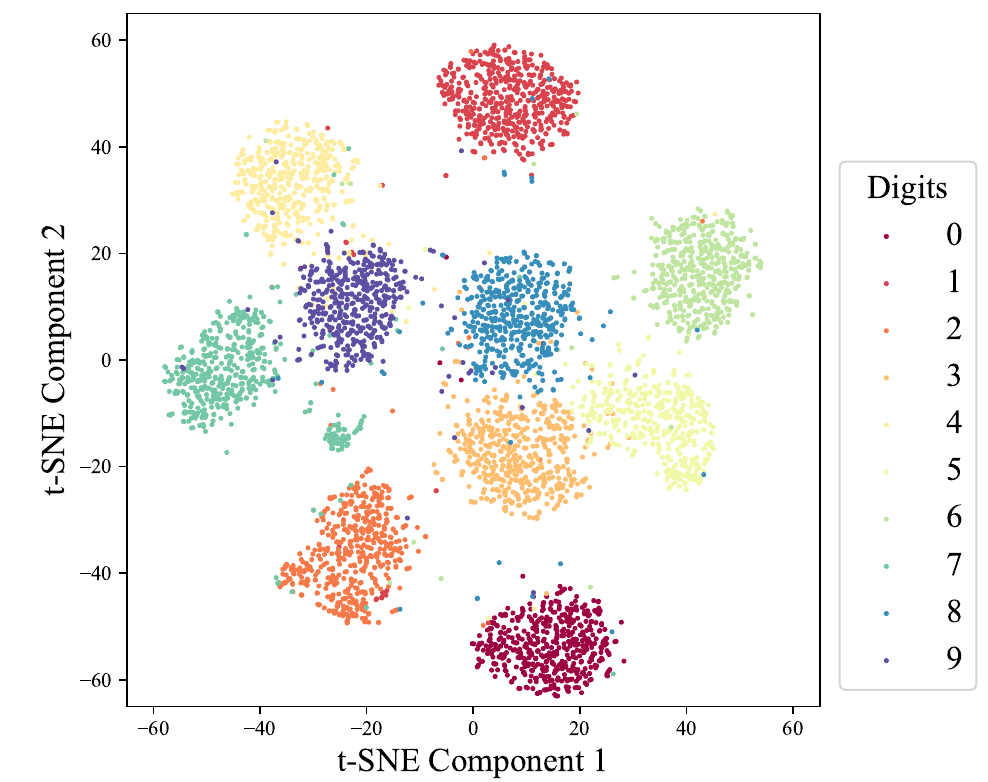}
    \caption{Geometric shape categories.\mbox{~~}}
    \label{fig:tsne_shape}
\end{subfigure}
\hfill
\begin{subfigure}[t]{0.25\linewidth}
    \includegraphics[width=\linewidth]{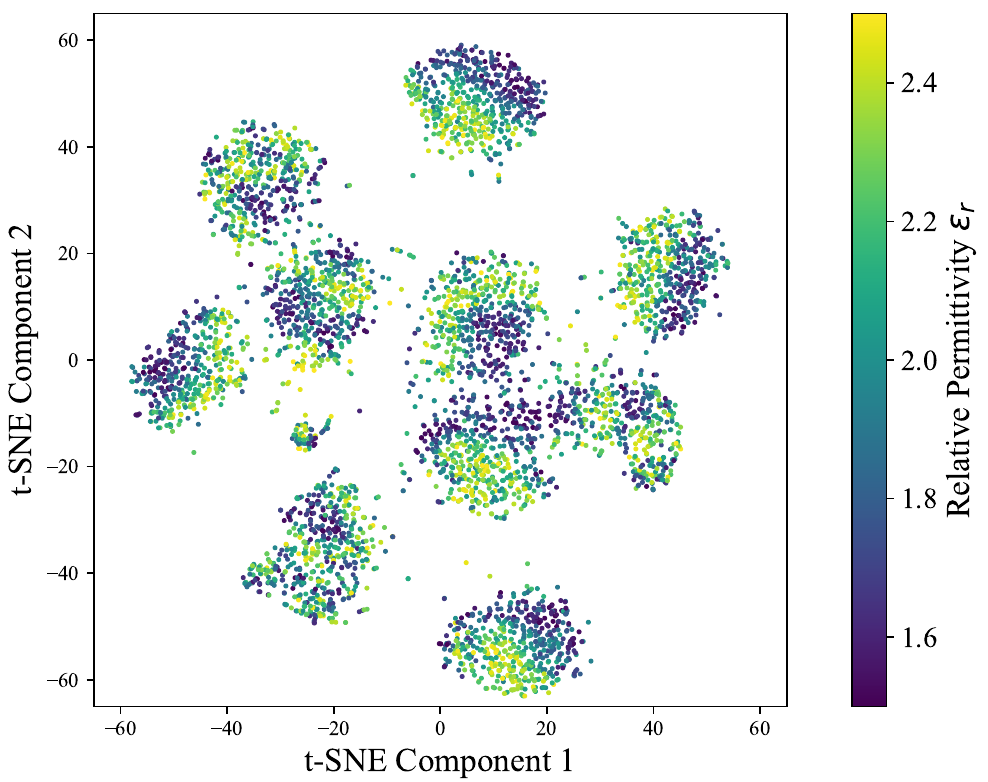}
    \caption{Relative permittivity.\mbox{~~~~~~}}
    \label{fig:tsne_epsr}
\end{subfigure}
\hfill 
\begin{subfigure}[t]{0.25\linewidth}
    \includegraphics[width=\linewidth]{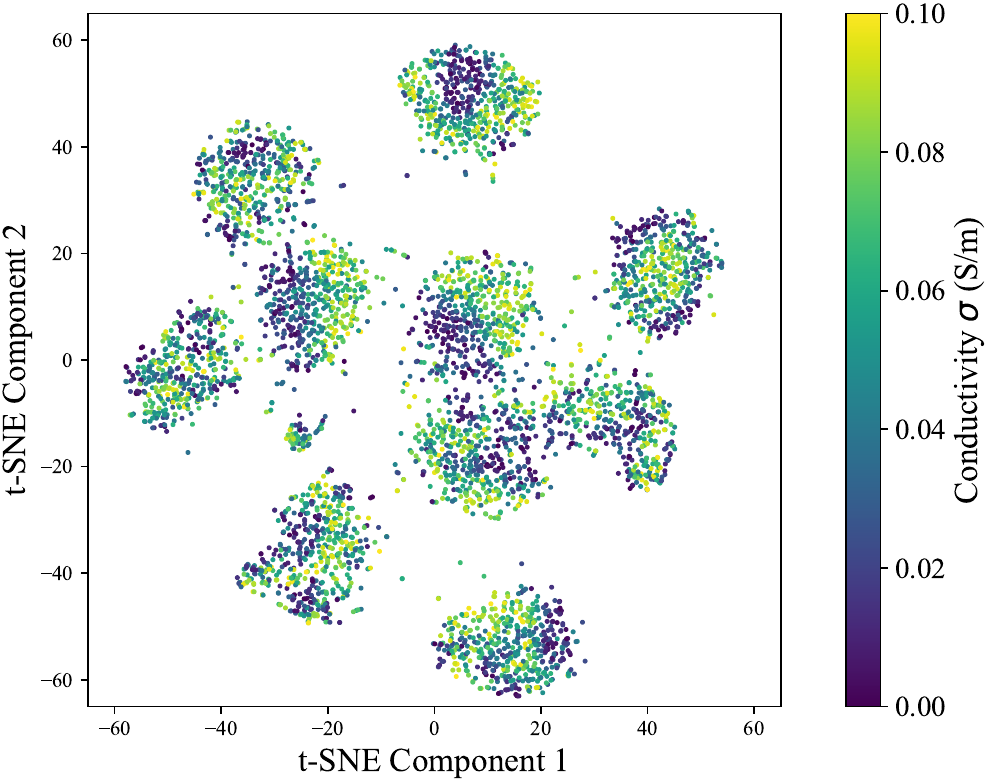}
    \caption{Conductivity.\mbox{~~~~~~~}}
    \label{fig:tsne_sigma}
\end{subfigure} 
\hspace{0.8cm}
\caption{t-SNE visualization of the latent space learned from the MNIST target dataset, where the multi-view channel encoder is implemented using IVT. }
\label{fig:t-SNE}
\vspace{-0.3cm}
\end{figure*}

\begin{figure*} [!t] 
\centering
\includegraphics[width=0.7\linewidth]{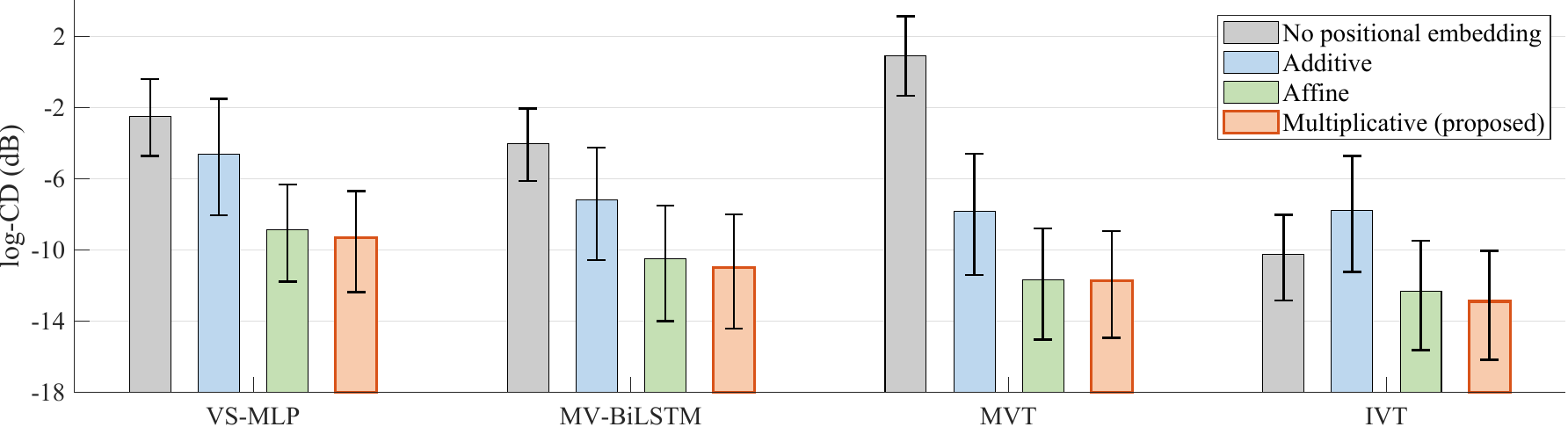}
\hspace{0.2 cm}
\caption{Performance of four multi-view channel encoders under different positional embedding schemes, where the quartiles Q1 and Q3 are marked.}
\label{fig:log_cd_bar_4pos_4algo}
\vspace{-0.3 cm}
\end{figure*}

Fig. \ref{fig:performance_signalSNRdB_Ns32_4algo} shows the test performance of the Gen-MV sensing models with four encoders across different SNRs, with the number of pilot symbols fixed at $L = 32$. 
It can be observed that as SNR increases, the performance of all four models gradually improves until stabilizing. 
Among them, MVT appears more susceptible to noise, which can be attributed to its smaller model inductive bias. 
Although IVT is also based on the attention mechanism, it fully exploits the structured nature of multi-view channel data, 
thereby achieving the highest performance upper bound while demonstrating optimal robustness against noise.

Fig. \ref{fig:performance_Nsymbol_differentSNR_IVT} evaluates the impact of the number of pilot symbols $L$ on the model's sensing performance under different SNRs. 
Experimental results indicate that although the model performance is limited under low SNRs, increasing the number of pilots progressively improves sensing accuracy, 
and lower SNRs require more pilots to achieve satisfactory sensing performance. 
Therefore, for low-SNR cases, increasing the number of transmitted pilot symbols can enhance channel estimation accuracy and subsequently improve target reconstruction quality, 
which is feasible for the static target sensing task considered in this work. 


To test the adaptability of the Gen-MV sensing method to environmental clutter, we randomly generate several circular clutter scatterers with a diameter of 0.05 m in an outer region ($0.5<|x|<1$, $0.5<|y|<1$).  
The EM property ranges of these clutter scatterers were consistent with those of the targets in the RoI. 
The model is trained under conditions with 1 -- 3 clutter scatterers. 
Fig. \ref{fig:performance_clutterN1to4_signalSNRdB_Ns32} illustrates the test performance of the IVT-based Gen-MV sensing model under different SNRs for scenarios containing different numbers of clutter scatterers.
The experimental results demonstrate that even with interference from environmental clutter, the proposed model can effectively extract target information within the RoI from multi-view CSI to achieve reconstruction, 
and it performs better with lower noise and fewer clutter scatterers. 
Moreover, the model remains functional when facing higher noise levels or stronger clutter interference than encountered during training. 

\vspace{-0.02cm}

The experimental results fully validate the applicability of the proposed scheme under non-ideal channel conditions. 
It is worth mentioning that our work presents a CSI-centric multi-view intelligent sensing framework, which does not rely on specific channel acquisition methods. 
In principle, employing more advanced channel estimation and pilot design schemes are expected to yield more accurate CSI, which in turn enhances the sensing performance of the proposed model.

\begin{figure*} [!t] 
\centering
\includegraphics[width=0.7\linewidth]{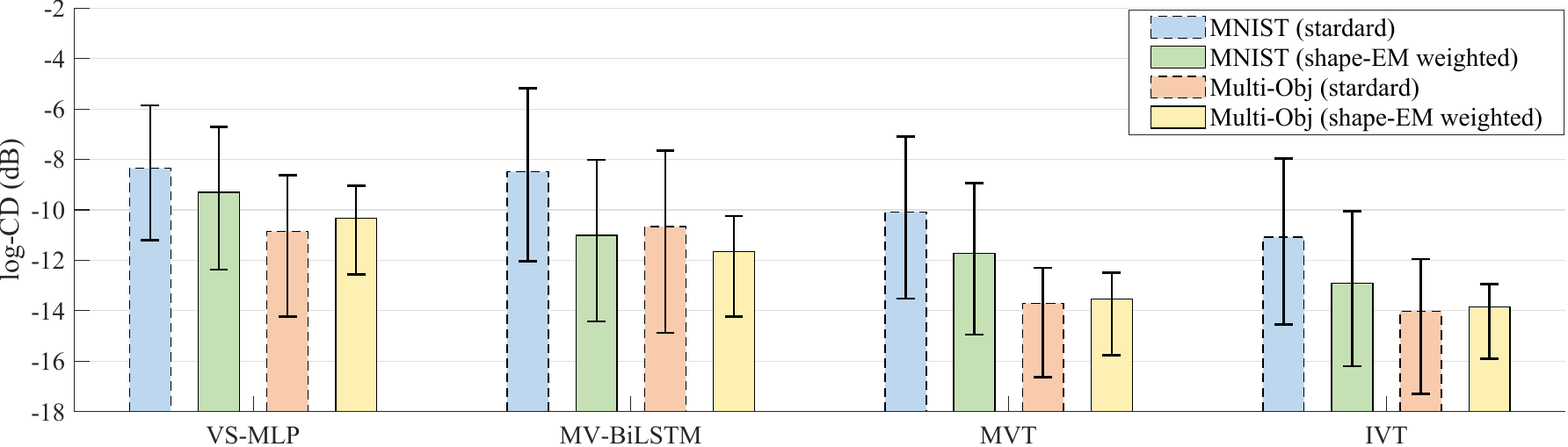}
\hspace{0.2cm}
\caption{Performance of four multi-view channel encoders on the MNIST and Multi-Obj datasets, with standard ($\gamma_{\mathrm{s}}=\gamma_{\mathrm{EM}}=0.25$) and shape-EM weighted ($\gamma_{\mathrm{s}}=0.45$, $\gamma_{\mathrm{EM}}=0.05$) diffusion loss, where the quartiles Q1 and Q3 are marked.}
\label{fig:log_cd_bar_2dataset_wo_shapeEM_4algo}
\vspace{-0.2 cm}
\end{figure*}

\begin{figure*}[!t]
    \centering
    \begin{subfigure}[t]{\linewidth}
    \centering
    \includegraphics[width=0.65\linewidth]{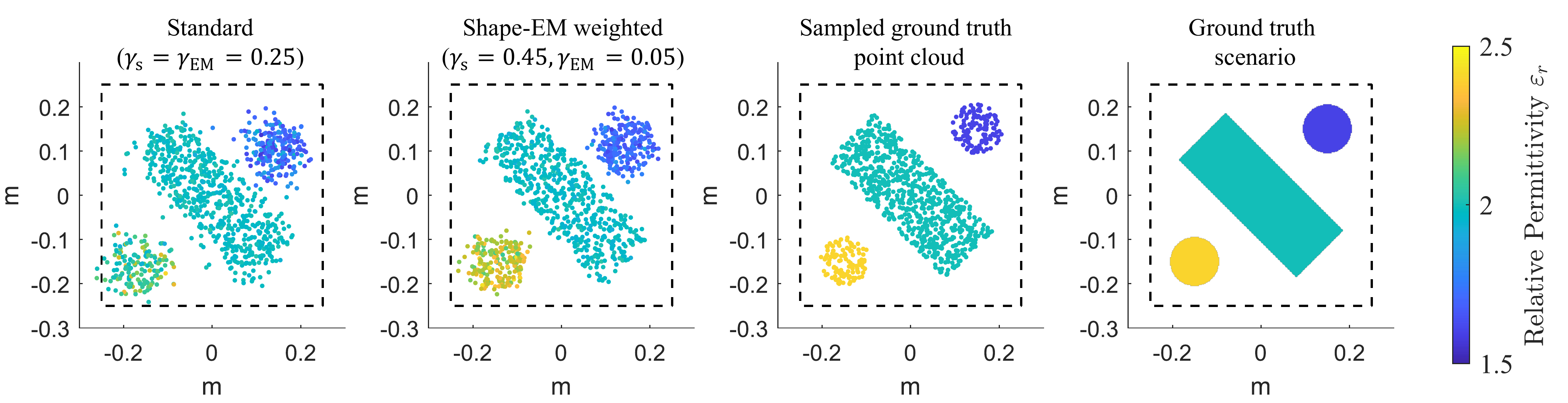}
    \vspace{-0.15cm}
    \caption{Relative permittivity.}
    \label{fig:example_epsr_wo_shape_EM}
    \end{subfigure}

    \vspace{0.1cm} 
    \begin{subfigure}[t]{\linewidth}
    \centering
    \includegraphics[width=0.65\linewidth]{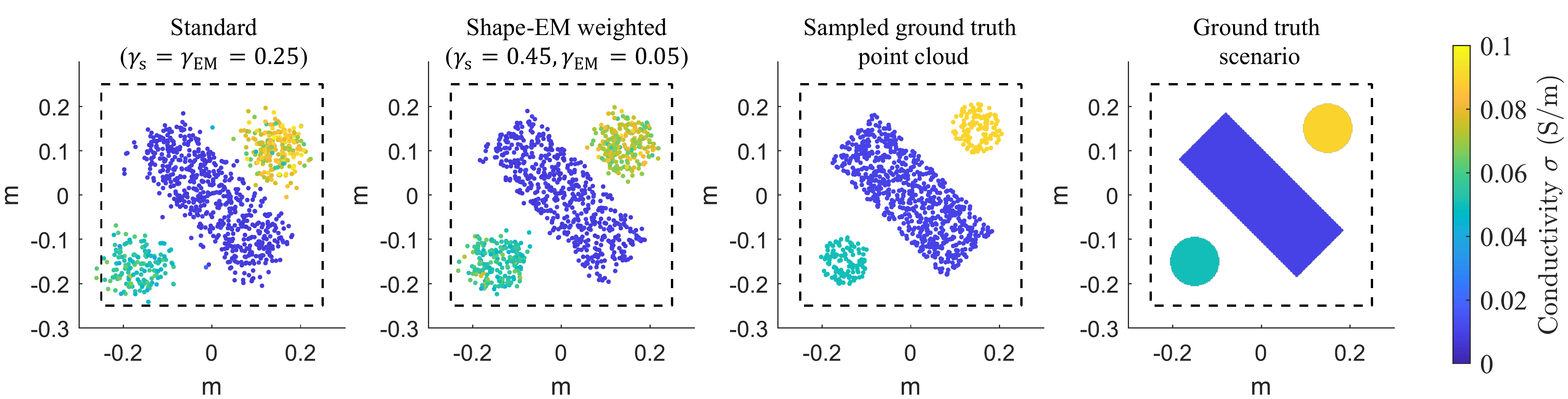}
    \vspace{-0.15cm}
    \caption{Conductivity.}
    \label{fig:example_sigma_wo_shape_EM}
    \end{subfigure}
    
\caption{Reconstruction results of IVT-based models using different loss weighting schemes on a hard sample from the Multi-Obj dataset. The RoI is marked with a dashed box.}
\label{fig:example_wo_shape-EM_weight}
\vspace{-0.3cm}
\end{figure*}

\subsubsection{Latent Space Representation of Target Features}
To analyze the distribution of latent representations of target features, we project the target latent code $\boldsymbol{z}$ generated by the multi-view channel encoder with IVT structure into a two-dimensional space using t-distributed stochastic neighbor embedding (t-SNE) \cite{van2008tSNE}. 
The visualization results of geometric shape categories, relative permittivity, and conductivity are shown in Fig. \ref{fig:t-SNE}. 
Due to the variational inference process of the Gen-MV framework and the constraints from the latent space prior model, 
the low-dimensional projections of the target latent codes exhibit clustering characteristics according to the geometric shape, with significant margins between most shape categories.
Within each cluster, the EM properties, including relative permittivity and conductivity, also exhibit smooth and regular distributions.
These phenomena reflect the structure and semantic information of the target latent space, which further indicates that our proposed channel encoder can effectively extract shape and EM property information of targets from multi-view channels, 
suggesting its potential applications for target classification and material detection.

\vspace{-0.2cm}
\subsection{Ablation Studies} \label{subsec:ablation_study}

\subsubsection{Ablation Study on Positional Embedding}
To validate the effectiveness of the proposed multiplicative positional embedding in processing channel data, we designed the following three comparative schemes:
(\romannumeral 1) No embedding of BS/UE positional information; 
(\romannumeral 2) Classical additive positional embedding, i.e., modifying (\ref{eq:pos_embed}) to $\boldsymbol{h}^{\mathrm{in}}_{b,u} = \boldsymbol{h}_{b,u} + \boldsymbol{\beta}(\boldsymbol{\xi}_{b,u})$; 
(\romannumeral 3) Affine-type positional embedding, i.e., modifying (\ref{eq:pos_embed}) to $\boldsymbol{h}^{\mathrm{in}}_{b,u} = \boldsymbol{\gamma}(\boldsymbol{\xi}_{b,u}) \odot \boldsymbol{h}_{b,u} + \boldsymbol{\beta}(\boldsymbol{\xi}_{b,u})$, 
where $\boldsymbol{\beta}$ and $\boldsymbol{\gamma}$ are fully connected layers. 
Fig. \ref{fig:log_cd_bar_4pos_4algo} illustrates the reconstruction performance of each encoder architecture under the four positional embedding schemes, with the first (Q1) and third (Q3) quartiles marked. 
Compared to the additive positional embedding in NLP, multiplicative positional embedding significantly enhances model performance, and this conclusion holds for all proposed multi-view channel encoder architectures. 
Furthermore, employing affine-type positional embedding (a combination of multiplicative and additive approaches) shows no performance gain over multiplicative embedding. 
These experimental results demonstrate the effectiveness and universality of the multiplicative operation in embedding BS/UE positions into channel features. 

Additionally, we notice that for the three baseline encoders, the absence of positional embedding renders the models nearly inoperative, but IVT still operates effectively under such conditions. 
This is because the interleaved operations in IVT implicitly capture the spatial correspondence between transmitters and receivers across multiple views, 
which enables the model to function even in the absence of explicit BS/UE location coordinates.  
Moreover, for IVT, explicitly embedding the view positions via the multiplicative operation remains more effective, whereas improper additive embedding can disrupt the model and degrade performance. 
These results further indicate that incorporating appropriate physical priors into the Gen-MV sensing model enhances its capability to extract environmental features. 

\subsubsection{Ablation Study on Shape-EM Weighted Diffusion Loss}

To validate the effectiveness of the proposed shape-EM weighted diffusion loss in improving reconstruction quality, 
we conduct experiments on the MNIST target dataset using the standard diffusion loss\footnote{Evidently, the standard diffusion loss is equivalent to using a shape-EM weighted diffusion loss with $\gamma_{\mathrm{s}}=\gamma_{\mathrm{EM}}=0.25$.} and the shape-EM weighted diffusion loss with coefficients $\gamma_{\mathrm{s}}=0.45$, $\gamma_{\mathrm{EM}}=0.05$. 
To further evaluate the proposed method for multi-object imaging, we refer to \cite{xing2025vbim, wang2023MSDLS} and construct a heterogeneous multi-target dataset, \textit{Multi-Obj}, to conduct the same ablation study, 
where several cuboids and cylinders with different EM materials are randomly placed within the RoI. 
For each object, the relative permittivity $\varepsilon_r$ is sampled from $[1.5, 2.5]$, and the conductivity $\sigma$ from $[0, 0.1]$ (S/m).
The evaluation results are presented in Fig. \ref{fig:log_cd_bar_2dataset_wo_shapeEM_4algo}. 

For the MNIST dataset, where the EM properties of the target in each scene are homogeneous, shape reconstruction is the main challenge. 
In this case, applying shape-EM weighted loss significantly improves overall reconstruction performance. 
In contrast, for the Multi-Obj dataset, the models have to estimate the position, shape, and EM properties of multiple targets simultaneously. 
Since the EM property distribution becomes complicated, placing more emphasis on shape reconstruction does not yield a clear improvement in average log-CD performance. 
However, according to error bars, the weighted scheme provides more consistent reconstruction performance across the dataset and performs better on hard samples compared to the standard diffusion loss. 
To provide an intuitive illustration, Fig. \ref{fig:example_wo_shape-EM_weight} shows the reconstruction results for a hard sample with compactly distributed objects in the Multi-Obj dataset, obtained from the IVT-based models trained with these two loss weighting schemes. 
It can be observed that even under heterogeneous multi-object conditions, assigning more weight to the shape component in the loss function still helps recover clearer object contours, which is beneficial for multi-object segmentation and localization. 
Therefore, the shape-EM weighted diffusion loss provides a flexible trade-off for joint shape and EM property sensing, 
and emphasizing shape reconstruction is practical in a wide range of scenarios.

\section{Conclusions}\label{sec:conclusion}
In this paper, we propose a novel conditional generative learning framework for multi-view wireless sensing, 
leveraging uplink CSI from multiple BSs and UEs in ISAC networks to achieve high-quality reconstruction of the target within the RoI. 
Overall, we develop an end-to-end sensing model by integrating multi-view channel fusion with controlled target generation, establishing a general Gen-MV sensing framework. 
In the proposed multi-view channel encoder, we design a multiplicative positional embedding to compensate for the impact of variable BS and UE positions on channel features. 
Subsequently, we implement multi-view fusion with classical architectures, and further propose an interleaved correlation learning scheme by incorporating physical information. 
The fused features guide the diffusion model to generate the target point cloud, and we introduce a weighted loss function to balance the spatial distribution differences of geometric shapes and EM properties of targets. 
The numerical results demonstrate that our proposed schemes can adapt to dynamic changes in quantities and positions of BSs and UEs, leveraging multi-view CSI to enhance environmental sensing. 
Ablation experiments further validate the efficacy of integrating physical priors into the GenAI framework. 

The core idea of this paper is to extract scenario information from multi-view CSI, which is not limited to the target EM imaging problem studied here, 
but also has the potential to be extended to a broader range of multi-view sensing and multi-device communication tasks, such as distributed radar sensing, multi-view joint channel estimation, 
and the bi-directional mapping between channels and environmental targets as preliminarily discussed in \cite{jiang2025bidirectionaldiffusion}.  
We hope our work can inspire future research in these emerging fields. 


\appendices{

{\color{black}
\section{BIM and BIM-CS} \label{app:BIM_BIM-CS}
According to (\ref{eq:discrete_total_field}) - (\ref{eq:y_hs}), we can formulate the multi-view channels at a single frequency $f_n$ as
\begin{equation} \label{eq:MV_spatial_channel_R3}
    \mathbf{H}_n = \mathbf{H}^{\mathrm{R-B}}_{n} \mathrm{diag}(\boldsymbol{\chi}_{n})\left[\mathbf{I} - \mathbf{G}_{n}\mathrm{diag}(\boldsymbol{\chi}_{n})\right]^{-1} \mathbf{H}^{\mathrm{U-R}}_{n}, 
\end{equation}
where $\mathbf{H}^{\mathrm{U-R}}_{n} = [\mathbf{h}^{\mathrm{U-R}}_{1,n}, \cdots, \mathbf{h}^{\mathrm{U-R}}_{U,n}], \; \mathbf{H}^{\mathrm{R-B}}_{n} = [\mathbf{H}^{\mathrm{R-B}}_{1,n}; \cdots; \allowbreak \mathbf{H}^{\mathrm{R-B}}_{B,n}]$. 
Equation (\ref{eq:MV_spatial_channel_R3}) can be vectorized as
\vspace{-0.05cm}
\begin{equation} \label{eq:vec_Hn_R3}
\vspace{-0.05cm}
    \mathrm{vec}(\mathbf{H}_n) = \mathbf{C}_{n} \boldsymbol{\chi}_{n}, 
\end{equation}
where
\vspace{-0.05cm}
\begin{equation} \label{eq:Cn_R3}
\vspace{-0.05cm}
    \mathbf{C}_{n} = \left((\mathbf{H}^{\mathrm{U-R}}_{n})^{\mathsf{T}}\left[\mathbf{I} - \mathbf{G}_{n}\mathrm{diag}(\boldsymbol{\chi}_{n})\right]^{-\mathsf{T}} \right) \ast \mathbf{H}^{\mathrm{R-B}}_{n}. 
\end{equation}
To facilitate the joint processing of multiple frequencies, we reformulate (\ref{eq:vec_Hn_R3}) into a real-imaginary separated form \cite{jiang2024EMpropertySensing}, \cite{jiang2025multi-BS}, 
\vspace{-0.05cm}
\begin{equation} \label{eq:SF_MV_real_linear_R3}
\vspace{-0.05cm}
    \mathbf{h}_{n} = \mathbf{A}_{n} \mathbf{x}, 
\end{equation}
where 
\vspace{-0.05cm}
\begin{equation} \label{eq:An_R3}
\vspace{-0.05cm}
    \mathbf{A}_{n} = \begin{bmatrix}
        \Re(\mathbf{C}_{n}) & -\frac{f_{c}}{f_{n}} \Im(\mathbf{C}_{n})\\
        \Im(\mathbf{C}_{n}) & \frac{f_{c}}{f_{n}} \Re(\mathbf{C}_{n})
    \end{bmatrix}, 
    \mathbf{h}_{n} = \begin{bmatrix}
        \Re\left(\mathrm{vec}(\mathbf{H}_n)\right)\\
        \Im\left(\mathrm{vec}(\mathbf{H}_n)\right)
    \end{bmatrix},
\end{equation}
and $\mathbf{x} = \left[\boldsymbol{\varepsilon}_r-1; \frac{\boldsymbol{\sigma}}{2\pi f_c \varepsilon_0}\right]$. 
By stacking the channel and measurement matrices across multiple frequencies, i.e. 
$\mathbf{h} = \left[\mathbf{h}_{1}; \cdots; \mathbf{h}_{N_c}\right]$, $\mathbf{A} = \left[\mathbf{A}_{1}; \cdots; \mathbf{A}_{N_c}\right]$, 
we obtain the multi-frequency multi-view channel model, 
\vspace{-0.1cm}
\begin{equation} \label{eq:MF_MV_real_linear_R3}
    \mathbf{h} = \mathbf{A} \mathbf{x}.
\end{equation}

Although (\ref{eq:MF_MV_real_linear_R3}) is formulated as a linear equation, 
the measurement matrix $\mathbf{A}$ and the quantity to be estimated $\mathbf{x}$ are correlated according to (\ref{eq:Cn_R3}) and (\ref{eq:An_R3}), 
making (\ref{eq:MF_MV_real_linear_R3}) essentially a nonlinear problem. 
Refer to \cite{jiang2024EMpropertySensing} and \cite{jiang2025multi-BS}, we first apply the first-order Born approximation to approximate $\mathbf{C}_n$ in (\ref{eq:Cn_R3}) by 
\begin{equation} \label{eq:BA_Cn_R3}
    \mathbf{C}_{n}^{(0)} = (\mathbf{H}^{\mathrm{U-R}}_{n})^{\mathsf{T}} \ast \mathbf{H}^{\mathrm{R-B}}_{n},  
\end{equation}
and the initial value $\mathbf{x}^{(0)}$ can be obtained according to (\ref{eq:SF_MV_real_linear_R3})-(\ref{eq:MF_MV_real_linear_R3}). 
Then, $N_{\mathrm{iter}}$ rounds of Born iterations are performed to alternately update the measurement matrix $\mathbf{A}$ and the variable $\mathbf{x}$ based on (\ref{eq:Cn_R3})-(\ref{eq:MF_MV_real_linear_R3}).
For the standard BIM \cite{wang1989BIM}, the initial solution and iterative update of $\mathbf{x}$ can be formulated as the following LS problem: 
\begin{equation} \label{eq:xi_R3}
    \mathbf{x}^{(i)} = \arg \min_{\mathbf{x}} \left\{\|\mathbf{h}-\mathbf{A}^{(i)}\mathbf{x}\|_2\right\}, \enspace
    \text{s.t.} \enspace \mathbf{0} \preceq \mathbf{x} \preceq \mathbf{\mathbf{x}}_{\mathrm{max}},
\end{equation}
where the superscript $i$ denotes the iteration count. 
We introduce prior knowledge of the maximum relative permittivity and conductivity to constrain the feasible region of $\mathbf{x}$, 
and use $\mathbf{\mathbf{x}}_{\mathrm{max}}$ to denote the corresponding upper bound vector. 
Furthermore, based on the sparsity of the RoI, we can reformulate (\ref{eq:xi_R3}) as the following compressed sensing problem \cite{jiang2024EMpropertySensing}, \cite{jiang2025multi-BS}, 
\begin{equation} \label{eq:update_x_R3}
\begin{split}
    &\mathbf{x}^{(i)} = \arg \min_{\mathbf{x}} \left\{\|\mathbf{h}-\mathbf{A}^{(i)}\mathbf{x}\|_2 + \lambda_{\mathrm{CS}} \cdot \|\mathbf{x}\|_{1,2} \right\}, \\
    &\text{s.t.} \enspace \mathbf{0} \preceq \mathbf{x} \preceq \mathbf{\mathbf{x}_{\mathrm{max}}},  
\end{split}
\end{equation}
where the $\ell_{1,2}$-norm regularization term is denoted as 
\begin{equation}
    \|\mathbf{x}\|_{1,2} = \sum_{d=1}^{D} \sqrt{\left(\boldsymbol{\varepsilon}_{r}[d]-1\right)^2 + \left(\frac{\boldsymbol{\sigma}[d]}{2\pi f_c \varepsilon_0}\right)^2},
\end{equation}
and $\lambda_{\mathrm{CS}}$ is the regularization coefficient. 
We refer to this algorithm as BIM-CS. 
Finally, the reconstructed images of the target's relative permittivity and conductivity are 
\begin{equation}
    \hat{\boldsymbol{\varepsilon}}_r = \mathbf{x}^{(N_{\mathrm{iter}})}[1:D] + 1, \;
    \hat{\boldsymbol{\sigma}} = 2\pi f_c \varepsilon_0 \cdot \mathbf{x}^{(N_{\mathrm{iter}})}[D+1:2D]. 
\end{equation}

Since both BIM and BIM-CS are pixel-based reconstruction algorithms, to enable comparison with the point cloud-based Gen-MV sensing models proposed in this paper, 
we apply K-means clustering with $K = 2$ to distinguish between target and background pixels. 
The reconstructed point cloud $\hat{\mathcal{X}}$ can be obtained by sampling the target region. 

}

}

\bibliographystyle{IEEEtran}
\bibliography{IEEEabrv.bib, myabrv.bib, ref.bib}


\end{document}